# Constrained CycleGAN for Effective Generation of Ultrasound Sector Images of Improved Spatial Resolution

**Xiaofei Sun[1], He Li[1] and Wei-Ning Lee[1,2*]**
[1]Department of Electrical and Electronic Engineering, The University of Hong Kong, Hong Kong
[2]Biomedical Engineering Programme, The University of Hong Kong, Hong Kong

E-mail: wnlee@eee.hku.hk

# Abstract

***Objective.*** A phased or a curvilinear array produces ultrasound (US) images with a sector field of view (FOV), which inherently exhibits spatially-varying image resolution with inferior quality in the far zone and towards the two sides azimuthally. Sector US images with improved spatial resolutions are favorable for accurate quantitative analysis of large and dynamic organs, such as the heart. Therefore, this study aims to translate US images with spatially-varying resolution to ones with less spatially-varying resolution. CycleGAN has been a prominent choice for unpaired medical image translation; however, it neither guarantees structural consistency nor preserves backscattering patterns between input and generated images for unpaired US images.
***Approach.*** To circumvent this limitation, we propose a constrained CycleGAN (CCycleGAN), which directly performs US image generation with unpaired images acquired by different ultrasound array probes. In addition to conventional adversarial and cycle-consistency losses of CycleGAN, CCycleGAN introduces an identical loss and a correlation coefficient loss based on intrinsic US backscattered signal properties to constrain structural consistency and backscattering patterns, respectively. Instead of post-processed B-mode images, CCycleGAN uses envelope data directly obtained from beamformed radio-frequency signals without any other non-linear postprocessing.
***Main Results***. *In vitro* phantom results demonstrate that CCycleGAN successfully generates images with improved spatial resolution as well as higher peak signal-to-noise ratio (PSNR) and structural similarity (SSIM) compared with benchmarks.
***Significance***. CCycleGAN-generated US images of the *in vivo* human beating heart further facilitate higher quality heart wall motion estimation than benchmarks-generated ones, particularly in deep regions.
The codes are available at https://github.com/xfsun99/CCycleGAN-TF2.

## 1. Introduction

Ultrasound imaging (US) is a non-invasive diagnostic tool used widely in clinical settings to assess internal organs, such as the heart, in real time. Single focus transmissions and dynamic receive beamforming with delay and sum are a fundamental image formation method implemented in commercial ultrasound systems and offers sufficient image quality for depicting tissue structures (Royer, 2019; Fenster and Downey, 1996). However, its performance may still be suboptimal for quantitative analysis of highly dynamic physiological events, such as blood flow, pulse waves, and tissue motion deep inside the human body. This is primarily due to the single focal spot at a time, which results in fine spatial resolution in a limited depth of field and a frame rate that typically ranges between 50-70 frames per second (fps) (Liebgott *et al.*, 2016). These two factors play a crucial role in speckle tracking, a technique used to quantify deformation and motion of biological tissues, such as the myocardium (Mondillo *et al.*, 2011). The spatial resolution and frame rate influence the accuracy and precision of speckle tracking, which, in turn, impact the overall assessment of cardiac function (D'hooge *et al.*, 2000).

Synthetic aperture imaging (Jensen *et al.*, 2006) and coherent plane wave compounding (Montaldo *et al.*, 2009) have been well established to enable two-way focusing in the entire field of view (FOV), resulting in improved spatial resolution, compared with standard single-focus or multi-focus transmissions; they achieve a uniform spatial resolution up to a certain imaging depth. In particular, coherent plane wave compounding (Montaldo *et al.*, 2009) is primarily realized in a linear array configuration; it achieves not only two-way focusing but also high frame rates. In a phased array configuration for cardiac ultrasound imaging, using diverging waves from virtual sources in a synthetic transmit fashion to achieve both two-way focusing and high frame rates has become prevalent (Papadacci *et al.*, 2014). This imaging method is known as diverging wave compounding but does not yield a uniform spatial resolution due to wave divergence as the wave travels to the deeper zone; spatial resolution worsens in a larger depth and towards two sides (i.e., at a larger azimuth angle) of the fan-shaped FOV. We hypothesize that such physical limitation may be tackled by deep learning (DL).

The goal of this study is thus to propose a DL method that translates phased or curvilinear-array images (with spatially-varying resolution) into quasi-linear array ones (with less spatially-varying resolution). More specifically, our method is based on Cycle-Consistent Generative Adversarial Network (CycleGAN) and incorporates ultrasound imaging physics into the model to yield US sector images with improved spatial resolutions comparable to linear-array ones.

In recent years, deep generative models, such as GAN (Goodfellow *et al.*, 2014) and diffusion models (Ho *et al.*, 2020) have become an increasingly prevalent strategy to learn complex features in the translation of different data domains with unprecedented success. GAN defines a minimax game in which a generator and a discriminator are two competing players. In the game, the generator is trained to learn an image mapping from the source domain to the target domain, and its output is a so-called generated image; the discriminator distinguishes the generated from the target image with a binary class label. A generative model is particularly appealing in the context of image-to-image translation. Although diffusion models have superior performance to GAN in natural image generation/translation, the disadvantages of diffusion models are presented in terms of low iterative sampling speeds and poor efficiency of model prediction (Yang *et al.*, 2022). GAN has demonstrated strong generalization ability in various data domains, particularly in medical imaging (Cai *et al.*, 2021; Ding *et al.*, 2021).

Current generic supervised-learning GAN-based models cannot address blind super resolution (SR) problems without low resolution (LR) and high resolution (HR) pairs. In clinical practice, that LR and HR ultrasound image pairs can hardly be acquired makes the supervised learning methods impractical. Besides, the difficulty in training a GAN was discussed in the literature (Radford *et al.*, 2015; Chartsias *et al.*, 2017; Martin Arjovsky and Bottou, 2017) that the generator may get stuck in a very narrow output distribution. This output distribution cannot represent a large variety of the real data distribution. Moreover, there is no concrete and interpretable metric yet for training progress. CycleGAN was consequently developed to overcome the drawbacks of individual GAN for improving the translation performance on natural images (Yuan *et al.*, 2018; Radford *et al.*, 2015).

For these GAN methods to apply to US image translation as proposed in our study, a substantial amount of paired linear-array and phased-array images will be required but hard to obtain in practice. Nehra et al. (Nehra *et al.*, 2022), Wolterink et al. (Wolterink *et al.*, 2017) and Chartsias et al. (Chartsias *et al.*, 2017) have used a CycleGAN (Zhu *et al.*, 2017) for medical image generation on unpaired data and achieved promising results. However, vanilla CycleGAN with only the adversarial loss and the cycle consistency loss can produce artifacts in the generated images and cannot ensure the structural consistency between the generated and input images without direct constraints between the two images as described in (Yang *et al.*, 2018;

Schaefferkoetter *et al.*, 2021). On the contrary, when the generated image is different from the input image, especially around the image edges, the physical tissue structure in the generated image is inconsistent with that in the input image with vanilla CycleGAN. To overcome this limitation, Zhang et al. (Zhang *et al.*, 2018) developed two CNNs, namely auxiliary segmentors, and introduced an extra loss to force the segmented ROI of the generated image to be identical to that of the input image. This network required training data with ground-truth ROI and further constrained training data requirements.

In addition to enforcing the structural consistency between the input and generated US images, preserving speckle pattern consistency is equally important. CycleGAN has alternatively been employed for ultrasound speckle reduction (Mishra *et al.*, 2018; Dietrichson *et al.*, 2018; Yi *et al.*, 2021). Speckle is an inherent feature in US imaging, and its statistical analysis has been used for tissue characterization. The similarity of speckle patterns reveals the degree of spatiotemporal correlation of tissue structures presented in US images (Cobbold, 2006).

To relax the requirement of paired data and tackle the aforementioned CycleGAN limitations, we hereby propose a constrained CycleGAN (CCycleGAN) with two extra loss functions (Section 2). More precisely, our method aims to generate US images with improved spatial resolution and better guarantee the structural and speckle pattern consistencies between the input and generated US images by adding an identical loss (Zhu *et al.*, 2017) and a correlation coefficient loss (Ge *et al.*, 2019). Our method is evaluated in aspects of spatial resolution and speckle statistics on an *in vitro* tissue-mimicking phantom (Section 3) and speckle tracking of *in vivo* US images of the beating human heart (Section 4).

## 2. Methods

The main objective of this study is to learn a non-linear mapping using unpaired US phased-array and linear-array images in an unsupervised way to further generate quasi-linear phased-array images that exhibit improved spatial resolution than the original US phased-array ones. It is regarded as an image style transfer between two different data domains. A CCycleGAN model is proposed and trained. **Fig. 1** shows the overall architecture of the proposed CCycleGAN.

### *2.1 Constrained cycle-consistent GAN (CCycleGAN)*

Our proposed CCycleGAN framework (**Fig. 1**) contains two generators (i.e., $G_A$ and $G_B$) and two discriminators (i.e., $D_A$ and $D_B$). $G_A$ and $G_B$ provide phased-to-linear and linear-to-phased mappings, respectively. $D_A$ is used to distinguish between original and model-generated phased array images and the corresponding discriminator $D_B$ is for the linear array counterparts.

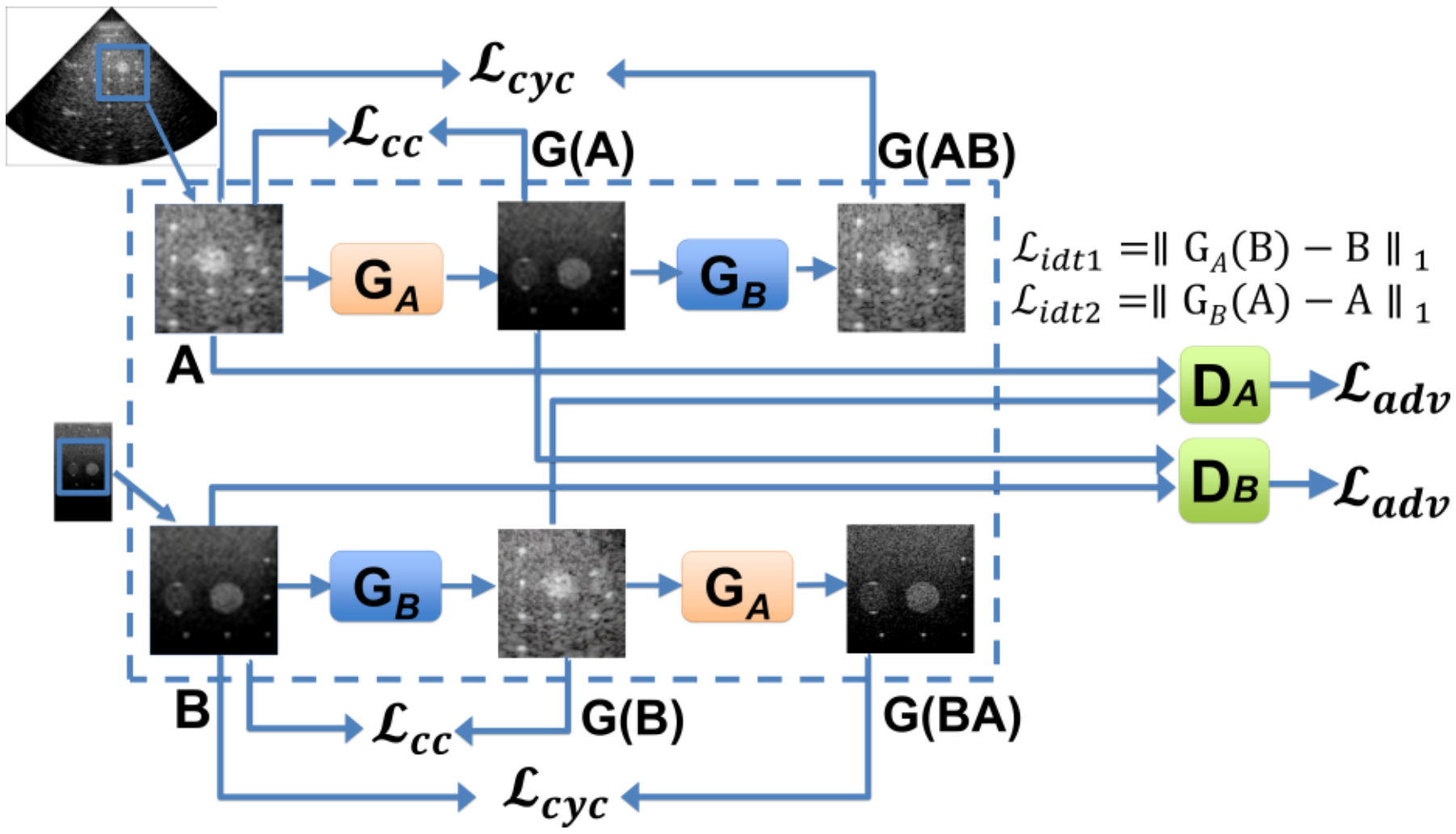

**Fig. 1.** The proposed constrained CycleGAN (CCycleGAN) framework. The cropped input images from a phased array and a linear array are merely exemplary images. The trained network is applied to the entire imaging field of view of the phased-array and linear-array US images. Two generators (i.e., $G_A$ and $G_B$) learn cross-domain mappings between linear-array and phased-array configurations. $G_A$ and $G_B$ provide phased-to-linear and linear-to-phased mappings, respectively. These mappings are constrained by adversarial, cycle-consistency ($\mathcal{L}_{cyc}$) and constrained-consistency losses. A and B denote the unpaired input phased-array and linear-array ultrasound images. $D_A$ and $D_B$ are discriminators. $\mathcal{L}_{cc}$ is correlation coefficient loss. $\mathcal{L}_{idt1}$ and $\mathcal{L}_{idt2}$ are identical loss functions. Note that italic *A* and *B* denote mapping functions and that A and B in the regular font style represent US images.

#### *2.1.1 Training loss*

The complete training loss includes an adversarial loss (Roy *et al.*, 2017), a cycle-consistency loss (Zhu *et al.*, 2017), and two newly proposed constrained-consistency losses, which are an identical loss and a correlation coefficient loss for preserving the continuity of structural information and inherent speckle patterns, respectively, in sequentially acquired US images of biological tissues whether static or dynamic.

***Adversarial loss***

The adversarial loss matches the intensity distribution of generated quasi-linear-array images to that of the target linear-array domain. The adversarial loss (Roy *et al.*, 2017) is applied to both the generator and discriminator. For the generator $G_A$ and its discriminator $D_B$, the adversarial loss is formulated as

$$\mathcal{L}_{adv}(\mathrm{G}_A, D_B, \mathrm{A}, \mathrm{B}) = \mathrm{E}_{B\sim P_{data}(B)}[log D_B(B)] + \mathrm{E}_{A\sim P_{data}(A)}[\log(1 - D_B(\mathrm{G}(\mathrm{A})))], \quad (1)$$

where A and B denote the unpaired input phased-array and linear-array ultrasound images, respectively, $E$ is the expectation, and $A\sim P_{data}$(A) and $B\sim P_{data}$(B) represent intensity distributions of A and B, respectively. Based on the game theory, during training, $\mathrm{G}_A$ tries to generate a quasi-linear-array image G(A) close to a real linear-array image, i.e., minG $\mathcal{L}_{adv}(\mathrm{G}_A, \mathrm{D}_B, \mathrm{A}, \mathrm{B})$, whereas $D_B$ is to distinguish between G(A) and B, i.e., maxG $\mathcal{L}_{adv}(\mathrm{G}_A, \mathrm{D}_B, \mathrm{A}, \mathrm{B})$.

Similarly, the adversarial loss for $G_B$ and $D_A$ is defined as

$$\mathcal{L}_{adv}(\mathrm{G}_B, \mathrm{D}_A, \mathrm{B}, \mathrm{A}) = \mathrm{E}_{\mathrm{A}\sim P_{data}(A)}[log \mathrm{D}_A(A)] + \mathrm{E}_{B\sim P_{data}(B)}[\log(1 - \mathrm{D}_A(\mathrm{G}(\mathrm{B})))], \quad (2)$$

***Cycle-consistency loss***

To prevent the generators from generating images that are uncorrelated with the inputs, a cycle-consistency loss (Zhu *et al.*, 2017) is adopted to force the reconstructed images G(AB) and G(BA) to approach their inputs A and B, respectively. This loss is written as

$$\mathcal{L}_{cyc}(\mathrm{G}_A, \mathrm{G}_B) = \mathrm{E}_{\mathrm{A}\sim P_{data}(\mathrm{A})}[\| \mathrm{G}(\mathrm{AB}) - \mathrm{A} \|_1] + \mathrm{E}_{B\sim P_{data}(\mathrm{B})}[\| \mathrm{G}(\mathrm{BA}) - \mathrm{B} \|_1], \quad (3)$$

where $\|\cdot\|_1$ is the $l_1$-norm. Then, the cycle-consistency loss imposes that $G_A$ and $G_B$ should be an inverse of each other, namely,$\mathrm{G}_B(\mathrm{G}(\mathrm{A})) = \mathrm{G}(\mathrm{AB}) \approx \mathrm{A}$ and $\mathrm{G}_A(\mathrm{G}(\mathrm{B})) = \mathrm{G}(\mathrm{BA}) \approx \mathrm{B}$.

***Constrained-consistency loss***

CycleGAN mainly utilizes unpaired training images and relies on the cycle-consistency loss to avoid the mismatches that occur in unsupervised-learning training. For image translation tasks, the classical loss function in conventional CycleGAN often results in blunt or coarsely-resolved structures (Wang *et al.*, 2018; Johnson *et al.*, 2016).

When the generated quasi-linear US image corresponds to the input phased-array image, the cycle consistency is well preserved, and G(AB) resembles the original input A. However, $\mathrm{G}_B(\mathrm{A})$ and $\mathrm{G}_A(\mathrm{B})$ may produce an inconsistent mapping of tissue structure. **Fig. 2** illustrates this problem. It was first observed that $G_A$ outputted a different phantom structure (**Fig. 2b**) from its input (**Fig. 2a**). It is thus critical that the framework can train the generators without altering physical tissue structure details in the input phased-array US images or input linear-array images. To keep structure consistency in any domain data themselves, Zhang et al. (Zhang *et al.*, 2018) introduced an identity loss to force the segmented ROI of the generated CT images to be identical to the ground-truth ROI of the input image. This loss was applied to CT image translation and denoising as a constrained function.

***Identical loss***

Inspired by (Zhang *et al.*, 2018), we adopt an identical loss function to constrain the generated US images through generator $G_A$ to preserve tissue structure details in the input phased-array US images. The generator $G_B$ should not change the physical structure depicted in the input linear-array images, either. To make $G_A$ and $G_B$ satisfy these conditions, the identical loss is defined as

$$\mathcal{L}_{idt}(G_A, G_B) = E_{B\sim P_{data}(B)}[\| G_A(B) - B \|_1] + E_{A\sim P_{data}(A)}[\| G_B(A) - A \|_1]. \quad (4)$$

The identical loss function can alleviate the drawback of cycle consistency, which does not enforce the structural similarity between the input and generated US images successively acquired in a cine-loop. Specifically, equation (4) constrains the mapping $G_A(B)$ from B to A via $G_A$, followed by backward mapping $G_B(A)$ from A to B via $G_B$ being similar to the original input, and vice versa. After applying the backward generator $G_B$, the phantom structure in the resultant image G(AB) became consistent with that in the original input A (**Fig. 2c**). Note that this identical loss is similar to the identical loss that prevents the generated image from transferring many texture features of input images (Zhu *et al.*, 2017).

***Correlation coefficient loss***

In addition to structural information, speckle patterns should also be preserved to ensure continuity of backscattering interferences among consecutive US images for tracing temporal evolution of speckle features. In conventional CycleGAN, perfect G(A) should be the image B, which is the linear-array counterpart of image A. However, neither cycle consistency nor identical loss can constrain the speckle features in successive images that are correlated. Another main challenge is that with unpaired images as the training data, we do not have the ground truth G(A) to compare with. To overcome this limitation, a correlation coefficient loss is proposed to explicitly enforce high backscattering (or speckle) similarity between the input A and its generated result G(A).

Specifically, the correlation coefficient, one of the most common multi-modality image registration metrics to enforce the constraint between the images, is hereby defined in (5) to restrain the speckle pattern deviation of the generated image G(A) from the input A. The same constraint is applied to G(B) and B.

$$\mathcal{L}_{cc}(G_A, G_B) = \frac{Cov(G(A),A)}{\sigma_{G(A)}\sigma_A} + \frac{Cov(G(B),B)}{\sigma_{G(B)}\sigma_B}, \quad (5)$$

where *Cov* denotes covariance, and σ represents variance.

***Complete training objective loss***

The complete training objective loss is formulated as

$$\mathcal{L}(G_A, G_B, D_A, D_B) = \mathcal{L}_{adv}(G_A, D_B) + \mathcal{L}_{adv}(G_B, D_A) + \lambda_1\mathcal{L}_{cyc}(G_A, G_B) + \lambda_2\mathcal{L}_{idt}(G_A, G_B) + \lambda_3\mathcal{L}_{cc}(G_A, G_B), \quad (6)$$

where $\lambda_1$, $\lambda_2$, and $\lambda_3$ control the weightings of the loss terms. In training, the log-likelihood objective in $\mathcal{L}_{adv}$ is replaced by a least-squares loss for training efficiency. To minimize $\mathcal{L}$, $D_A$ and $D_B$ update their weights when the parameters of $G_A$ and $G_B$ are fixed. Similarly, $G_A$ and $G_B$ are updated while keeping parameters of $D_A$ and $D_B$ fixed.

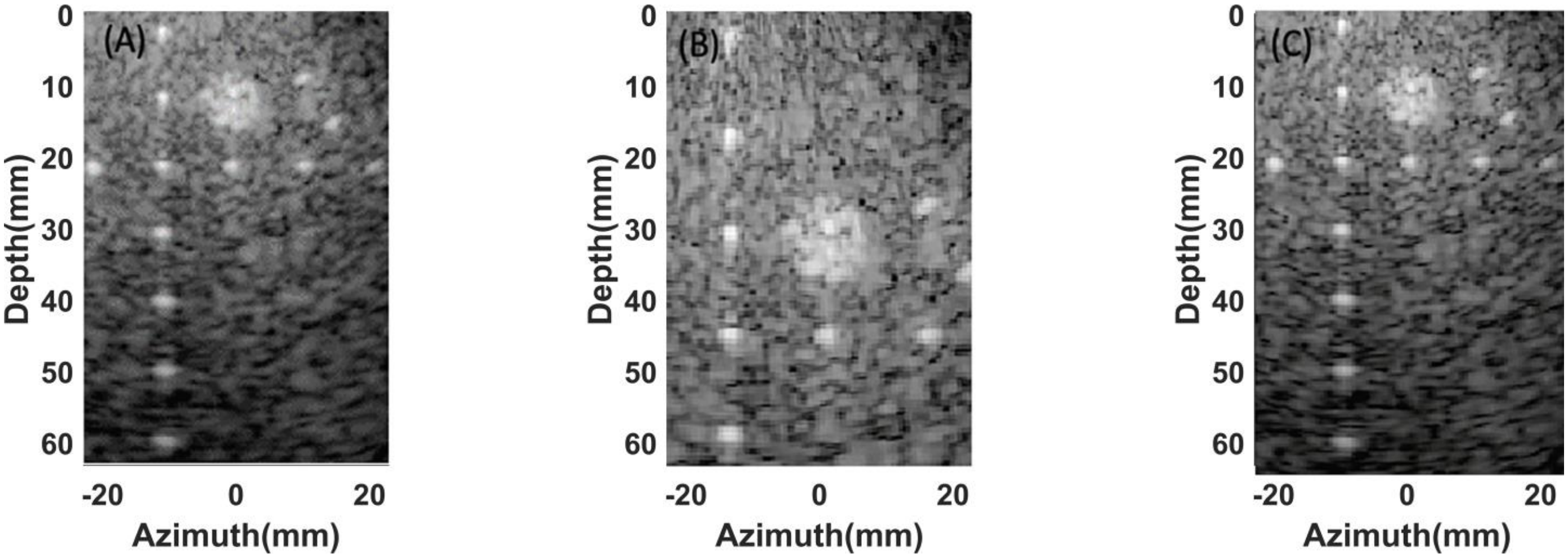

**Fig. 2.** Exemplary ultrasound images of a commercial multi-purpose tissue-mimicking phantom showing the effectiveness of the proposed identical loss: (A) region of interest (ROI) in an original phased-array image as the input image, (B) a model-generated quasi-linear-array image by $G_A$ without the proposed identical loss. (C) a model-generated quasi-linear-array image by $G_A$ with the proposed identical loss. This demonstrated that the structural consistency was preserved in (C) but not in (B).

## 2.2 Network architectures

### *2.2.1 Generators*

With the proposed constrained-consistency losses, our CCycleGAN architecture is developed and optimized to achieve improved spatial resolution in the sector FOV. The architecture of our generators is composed of four trainable neural networks as shown in **Fig. 3a.** We use the network structure described in (Kang *et al.*, 2019). The first convolution layer consists of 256 sets of 3 × 3 kernels, which are of the smallest possible kernel size, give 8-neighborhood pixel information, and are computationally efficient to produce 256 channel feature maps after convolution operations. To reduce network complexity while keeping accuracy, we only deploy five modules. Each module is composed of 1) three sets of convolutions, batch normalization, and ReLU and 2) one skip connection with a ReLU. Convolution layers in each module use 256 sets of 3×3×256 kernels. Different from the original network structure, our proposed generator network reserves a concatenation layer (**Fig. 3b**) that efficiently concatenates the input of each module and enhances the output of the last module (i.e., Module 5 in **Fig. 3a**). The last module is further followed by a convolution layer with 256 sets of 3×3×1536 kernels. The concatenation layer as shown in **Fig. 3b** has a boosting effect that combines multiple weak inputs to provide a strong output (Schapire *et al.*, 1998). That perfect generation or translation satisfies for all the cascades of encoder-decoder networks is assumed. Furthermore, the generation condition for the generator network up to $N$-layer can be described by a matrix-vector multiplication as follows:

$$\begin{aligned} L_1 &= c^{(1)} \otimes d^{(1)} \\ &\vdots \\ L_N &= c^{(N)} \otimes d^{(N)} \otimes d^{(N-1)} \dots \otimes d^{(1)}, \end{aligned} \tag{7}$$

$$c^{(i)} = \begin{cases} \left(c^{(i-1)} \otimes e^{(i)}\right), & i = 1, 2, \dots, N \\ f, & i = 0 \end{cases}, \tag{8}$$

where $L$ is a decoded matrix given input representation, which is denoted by $f$; $c$ is the convolution coefficient, $e$ and $d$ are respectively the encoder and the decoder, and the subscript $i$ denotes the $i$-th layer of the generator. For an intermediate encoder output $c^{(i)}|_{i=1}^{N}$, let $h^{(i)} = d^{(i)} \otimes d^{(i-1)} \ldots \otimes d^{(1)}$ , and a boosted decoder can be defined by combining multiple decoders as

$$L_i = \sum_{i=1}^{N} \omega_i \left(c^{(i)} \otimes h^{(i)}\right), \tag{9}$$

where $\omega_i$ is the weight, and $\sum_{i=1}^{N} \omega_i = 1$. This procedure can be realized with a single convolution after concatenating encoder outputs as shown in **Fig. 3b**.

The last convolution layer uses 3×3×256 convolution kernels. Finally, the generator has an end-to-end skip connection (He *et al.*, 2016) as shown in **Fig. 3c** to concatenate various levels of hidden features in the input data robustly and preserve the features of the input data as proposed in (He *et al.*, 2016; Ahn and Yim, 2020).

### *2.2.2 Discriminators*

The network architecture of discriminators $D_A$ and $D_B$ using blocks of Conv-InstanceNorm-LeakyReLU layers with 4 × 4 filters and a 2-stride is illustrated in **Fig. 4**. This architecture is based on PatchGAN (Isola *et al.*, 2017) with 70 × 70 receptive fields, which detects whether image patches are real or generated. Specifically, the discriminator in our proposed CCycleGAN does not use batch normalization but instead instance normalization with LeakyReLU for preventing the vanishing gradients (Ulyanov *et al.*, 2016). The discriminator consists of five convolution layers, including the last fully-connected layer. The first convolution layer uses 64 sets of 4 × 4 kernels, and the number of kernels in a subsequent layer doubles that of the preceding layer, except for the last fully-connected layer. After the last layer, 6 × 6 feature maps are obtained, and the $l_2$-loss is calculated. Owing to an advantage of PatchGAN, arbitrarily large input images can be applied to the discriminator by summing up the $l_2$-losses from all 64 × 64 patches, after which a final decision is made.

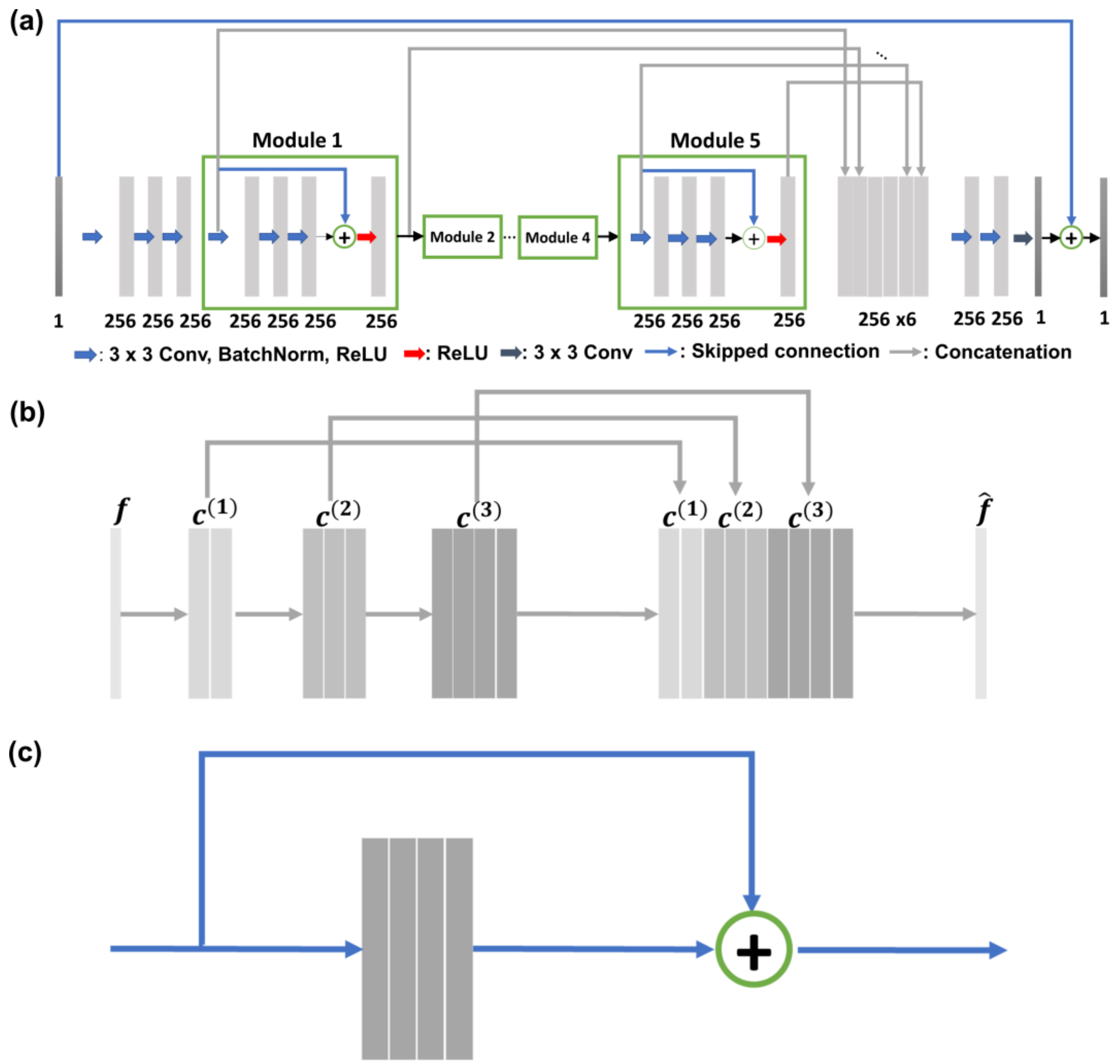

**Fig. 3.** (a) Network architecture of the generator in the proposed CCycleGAN;(b) Concatenation layer and (c) skip connection used in **Fig. 3a.**

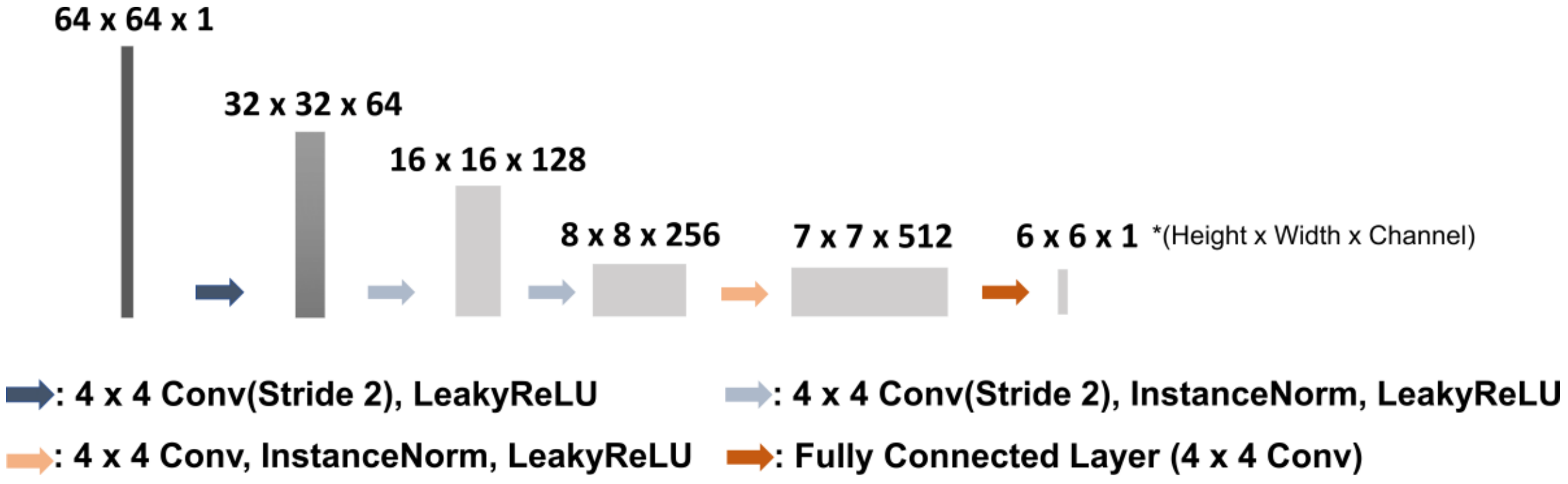

**Fig. 4.** Network architecture of the Patch-GAN discriminator in the proposed CCycleGAN

## 3. Datasets, experiments and quantitative analysis

### *3.1 Datasets*

The CCycleGAN framework was trained on an *in vitro* dataset and tested on *in vitro* and *in vivo* datasets. *In vitro data* were ultrasound radio-frequency (RF) data of a multi-purpose, multi-tissue ultrasound phantom (CIRS Model 040GSE) scanned in three different views as shown in **Fig. 5** using a Verasonics® Vantage 256 ultrasound system equipped with an L7-4 linear-array probe (5.2 MHz) and a Philips P4-2 phased-array probe (2.5 MHz). The acquisition sequence and image formation followed the ultrasound imaging method with cascaded dual-polarity waves, which is based on a unique designed coded excitation scheme, diverging waves, and a synthetic transmit aperture, to achieve high signal-to-noise ratio (SNR) and high frame-rate ultrasound imaging (Zhang *et al.*, 2017; Zhang *et al.*, 2019). This higher SNR is advantageous to cardiac ultrasound imaging. There were 1080 linear-array images (960 cases for training and 200 cases for validation; image depth: 60.0 mm) and 1320 phased-array images without scan conversion (1100 cases for training and 220 cases for validation; image depth: 112.1 mm).

In addition to internally collected *in vitro* data, a set of linear-array images from the public PICMUS challenge database (Liebgott *et al.*, 2016) were also included but only for training. The selected PICMUS data were also acquired using a Verasonics® system with a linear array probe operating at 6.25 MHz and consisted of 75 linear-array in-phase quadrature (IQ) data (image depth: 50.0 mm) of the same ultrasound phantom (CIRS Model 040GSE). All RF data from our lab and IQ data from PICMUS underwent envelop detection to yield envelope data whose sizes were standardized to 256-by-256 in units of samples. Note that the CCycleGAN framework in its present form took the envelope data as the input.

A total of 2135 US images, including 1035 (=960+75) linear-array US images and 1100 phased-array US images, were used as the training dataset, and 10% of them (213 cases) were used for parameter optimization. A total of 420 (i.e., 200 linear-array and 220 phased-array) US images were used as the validation dataset to evaluate the accuracy of our proposed CCycleGAN built in the training process while tuning model hyperparameters.

As one key feature of CCycleGAN, an unpaired dataset comprised of linear-array and phased-array envelop data was used to translate phased-array to quasi-linear array images. The test *in vitro* data were 200 phased-array US phantom images (image depth: 112.1 mm) in View 2 (the top right in **Fig. 5b**). *In vivo* ultrasound data of human beating hearts (image depth: 195.9 mm) were obtained by our Vantage 256 system with a P4-2 phased-array probe operating at 2.5 MHz. The human cardiac ultrasound data collection was approved by the Institutional Review Board at the University of Hong Kong (UW13-566), and each study subject provided written informed consent.

### *3.2 Deployment*

The Tensorflow 2.0 implementation of our CCycleGAN was employed. The training was performed on a computer with an Nvidia Quadro P6000 graphics processing unit (GPU). The implementation of our method took 200 epochs of learning. The training was constrained by minimizing the loss function (6) as listed in Section 5 with loss weights $\lambda_1$ = 10, $\lambda_2$ = 5, and $\lambda_3 = 5$. Adam (Kingma and Ba, 2014) optimization was used to train all networks. In the first 100 epochs, we set the learning rate at 0.0002 and linearly decreased it to zero over the next epochs. The size of the mini-batch was 10. A Gaussian distribution initialized convolutional kernels randomly. The generator and the discriminator were updated in every iteration. After the training ended, the neural network and the learned weights were exported as a graph and stored in a single TensorFlow checkpoint (ckpt) file. Only the generator was exported because the discriminator was required to be updated during the training. The model file was then loaded directly in Python. We normalized the intensities of the input phased-array and linear-array US images to the range [0, 1].

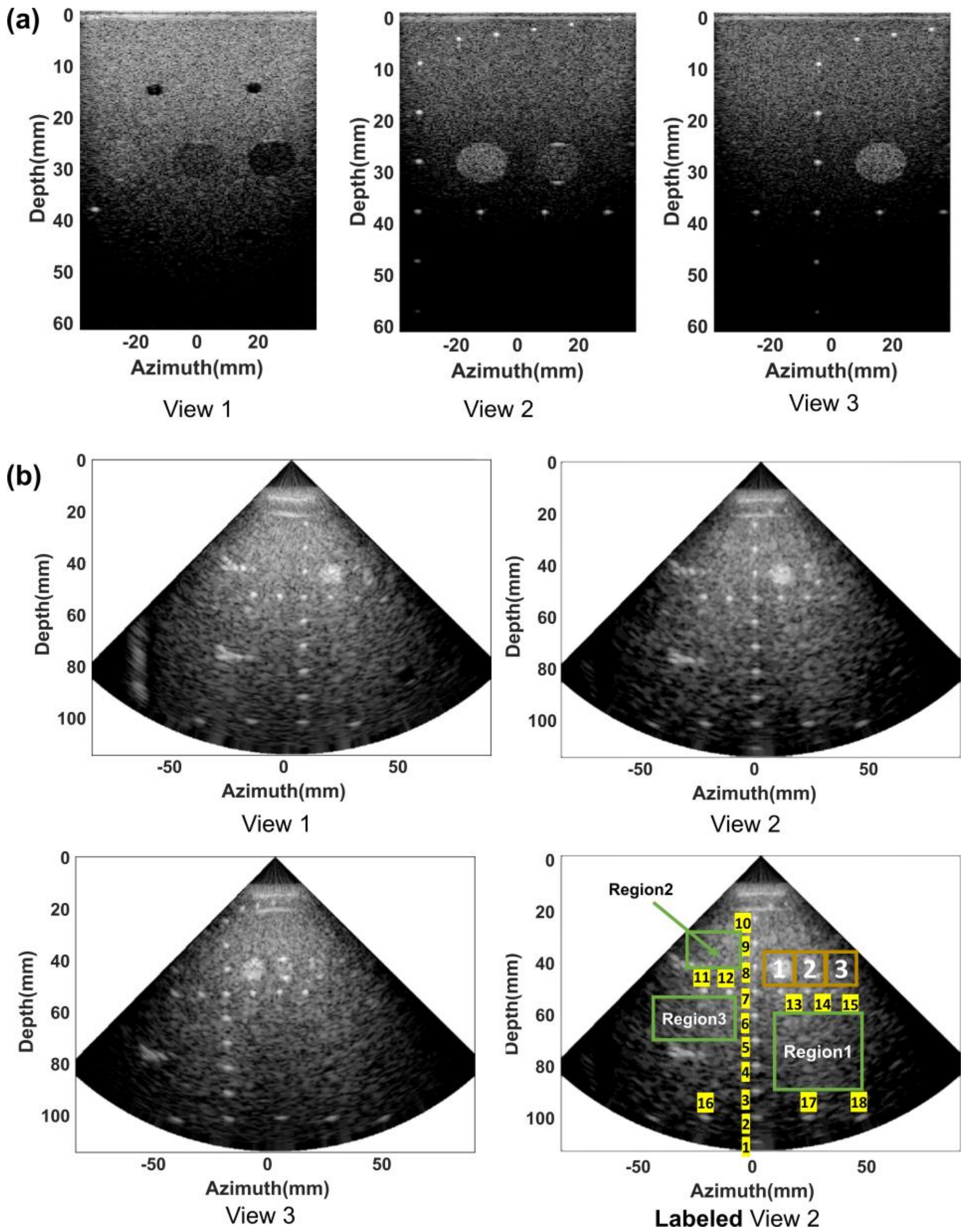

**Fig. 5** (a) Linear-array B-mode images of the same multi-purpose multi-tissue phantom (**Fig. 2**) taken at three different locations. (b) Phased-array B-mode images of the same phantom (**Figs. 2 and 5a**) in three different views. Points and regions of interest (ROIs) are numbered.

### *3.3 Quantitative analysis*

#### *3.3.1 Spatial resolution*

The spatial resolution of a 2D ultrasound image includes axial and lateral resolutions, which are quantified as the full width at half maximum (FWHM) beam widths in the axial and lateral directions, respectively, and typically in units of millimeter. FWHM is the full width of the point spread function (PSF) at half of its maximum value. In our study, FWHM is defined as the range of two coordinates (along axial or lateral direction) in a ROI of target point when the intensity of the ROI is equal to half of the peak intensity.

#### *3.3.2 Speckle statistics*

Speckle statistics were evaluated using the Nakagami statistical method described in (Yu *et al.*, 2015). This method estimates the shape parameter $m$, which depends on the statistical distribution of the envelope of the backscattered RF signals. When $m$ is equal to 1 or smaller than 1, the signal envelope follows a Rayleigh distribution or a pre-Rayleigh distribution, respectively.

#### *3.3.3 Speckle tracking*

As briefly mentioned in Section 3.1, this study also tested ultrasound images of the *in vivo* human heart to investigate whether DL-model-generated images could preserve the inter-frame correlation of a highly dynamic tissue or organ for reliable quantification of tissue/organ kinematics. The heart was imaged at a compounded rate of 400 fps in the apical two chamber (2CH), apical four chamber (4CH), and mitral-valve-level parasternal short-axis (SA) views. To quantify kinematics of the heart from ultrasound images, a cross-correlation-based speckle tracking method was employed (Li *et al.*, 2016; Li and Lee, 2017). Please note that there was no modification made to the adopted speckle tracking method for model-generated ultrasound images. The same algorithm parameters were used for all the original and model-generated images. For each cine loop of data as shown in **Fig. 6**, images within one complete cardiac cycle were selected according to synchronously acquired electrocardiogram (ECG). The cardiac cycle either started from the $R$ wave to the next $R$ wave or from the $T$ wave to the next $T$ wave of the ECG signal. To test the fidelity of model-generated cardiac US images for speckle tracking, 2CH, 4CH and SA views with 273, 260 and 305 cardiac US frames in one cycle, respectively, were included.

Displacements were estimated using a speckle tracking method (Li *et al.*, 2016; Li and Lee, 2017) that was previously validated on phantoms and *in vivo* datasets. To make a fair comparison, the same speckle tracking method using different deep learning model-generated US images were adopted to produce motion estimates. Please note that there was no modification made to the adopted speckle tracking method for model-generated ultrasound images. The same algorithm parameters were used for all the original and model-generated images. Structural similarity (SSIM) and root mean squared deviation (RMSD) were used as performance metrics of motion estimation. SSIM was calculated to evaluate the similarity between the displacement maps from the reference and those from deep learning models. RMSD was calculated between a pre-deformed and the motion-corrected post-deformed frame as an alternative way of quantifying motion estimation accuracy when the *in vivo* ground truth is unavailable. The post-deformed frame was corrected for its 2-D motion from the pre-deformed frame using a tracking-followed-by-recorrelation method (Li *et al.*, 2016) When motion estimation is perfect, a post-deformed frame after motion correction should be identical to the corresponding pre-deformed frame, and RMSD should be zero.

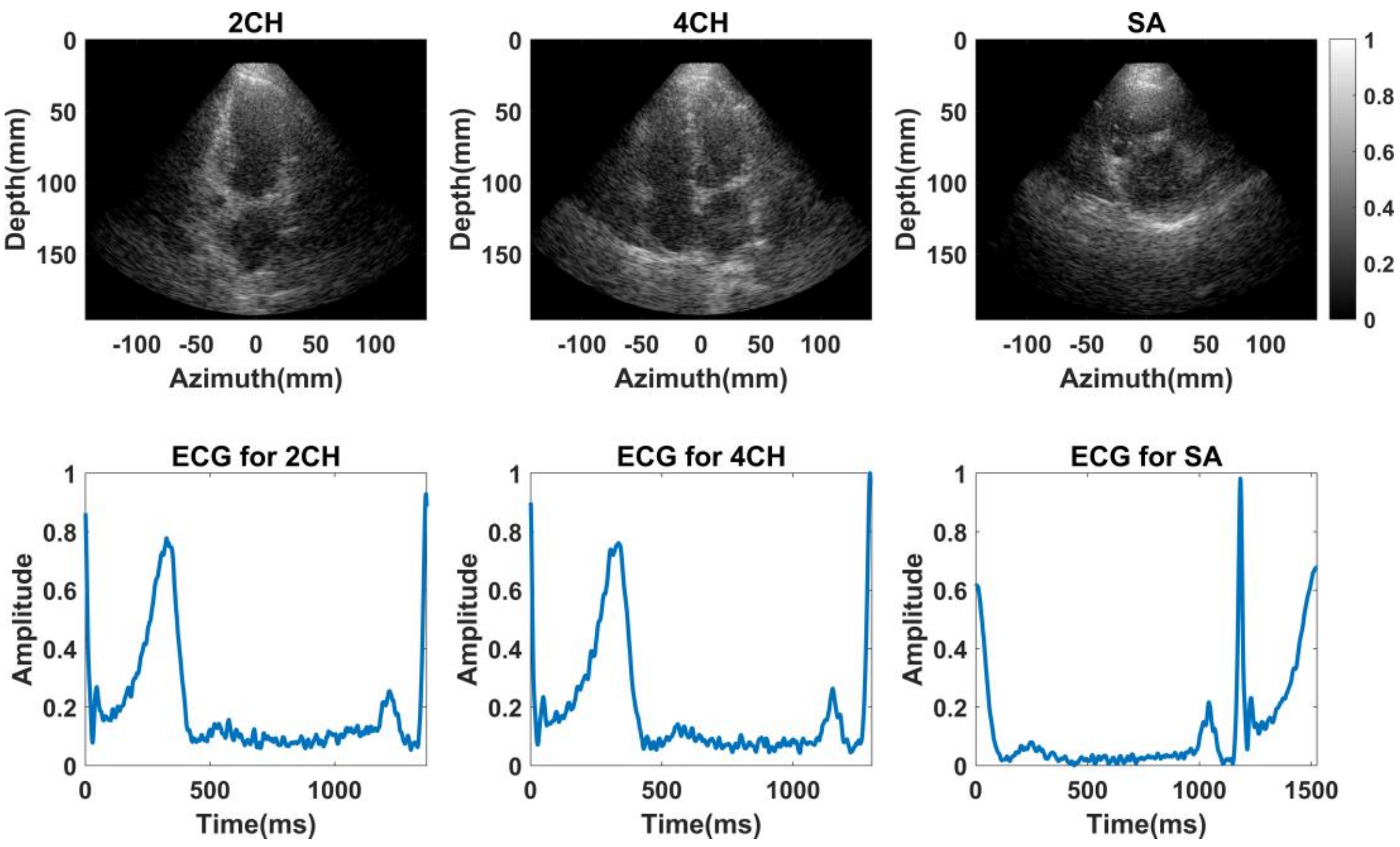

**Fig. 6.** **Top row**: exemplary cardiac US images from the left to the right are apical two chamber (2CH), apical four chamber (4CH), and parasternal short-axis (SA) views. **Bottom row**: the corresponding electrocardiograms (ECGs) of one complete cardiac cycle.

### *3.4 Ablation study*

To evaluate their individual effects on the model performance and compatibility with the existing adversarial and cycle-consistency losses, an ablation study was performed over the proposed identical and correlation coefficient losses incorporated into losses of CCycleGAN.

## 4. Results

Three state-of-the-art deep learning-based models, including Laplacian Pyramid Super-Resolution Network (LapSRN) (Tai *et al.*, 2017), Super Resolution (SRGAN) (Ledig *et al.*, 2017), and CycleGAN (Zhu *et al.*, 2017) were selected as benchmarks to evaluate model performance in aspects of spatial resolution, speckle statistics, quality of speckle tracking, SSIM, and PSNR. Besides, the effectiveness of the primary components of our generator network, i.e., identical loss and correlation coefficient loss, was shown. LapSRN is a CNN-based deep learning method for image generation with improved spatial resolution, whereas SRGAN and CycleGAN are GAN-based methods. As LapSRN and SRGAN were performed with paired images, it was necessary to create paired images for training. Inspired by the generation of high-resolution ultrasound images in (Liu *et al.*, 2021), one strategy was to utilize the network-based interpolation to up-sample the original phased-array images to obtain paired high-resolution images. The input paired images consisted of original phased-array and up-sampled phased-array images. CycleGAN and CCycleGAN used unpaired linear-array and phased-array images. These benchmark comparison networks were trained with a similar hyperparameter search.

### *4.1 Spatial resolution*

Labeled View 2 in **Fig. 5b** shows 18 numbered point targets that were used for spatial resolution analysis of 200 test US images of the phantom. **Fig. 7** shows original and deep learning model-generated phased-array images. Fig. 8 shows mean axial and lateral resolutions at point targets along Lines 1, 2, and 3 labeled on original phased-array images in labeled View 2 (**Fig. 5b**). CCycleGAN achieved better resolutions along Lines 1, 2, and 3 than the original image and benchmarks. The point target indicated by the orange arrow in the original image (**Fig. 7a**) was more pronounced in the model-generated images. The point target indicated by the blue arrow and the other points on the same horizontal Line 3 could be observed in the CCycleGAN-generated image but not in the CycleGAN-generated one. Similarly, the regions labeled with 3 and 4 were also enhanced in the CCycleGAN-generated image.

**Figs. 9a, 9c and 9e** summarized the mean value of the lateral resolution at each numbered point target in the original phased-array images and model-generated ones by our proposed model and benchmarks. CCycleGAN-generated phased-array images exhibited the best lateral resolutions among all. Our proposed CCycleGAN and CycleGAN improved the lateral resolution by on average 9.6% and 5.2% when compared with the original image, respectively. The lateral resolution worsened by 2.7% and 1.3% in cases of LapSRN and SRGAN, respectively. Similarly, from **Figs. 9b, 9d and 9f**, CCycleGAN-generated images presented comparable or better axial resolutions than the original phased-array images. On average, CycleGAN and CCycleGAN improved the axial resolution slightly by 3.2% and 1. 52%, respectively, while LapSRN, and SRGAN deteriorated the axial resolution by 1.6%and 0.2%, respectively. Most importantly, lateral resolutions at points 2-3 and 16-18, which were in the deep zone, were improved by CCycleGAN, not benchmarks.

### *4.2 Speckle statistics*

**Tables 1-3** summarized speckle statistics of the same set of 200 US phantom frames that were used for spatial resolution analysis. A paired student *t*-test statistic was used to evaluate if there was a significant difference in the mean statistical shape parameters, *m* values, between the original and model-generated images in the corresponding local regions, including Regions 1-3 (Labeled View 2 in Fig. 5b). As these regions were at different depths, the shape parameters *m* from these three regions were estimated separately. All the *p*-values were approximately $10^{-8}$, which was much smaller than the pre-defined significance level of 0.05 in the *t*-test, thus indicating that the original and model-generated images presented statistically different speckle patterns. Yet, compared with three benchmarks, CCycleGAN generated images with speckle statistics that were closer to those of the original images.

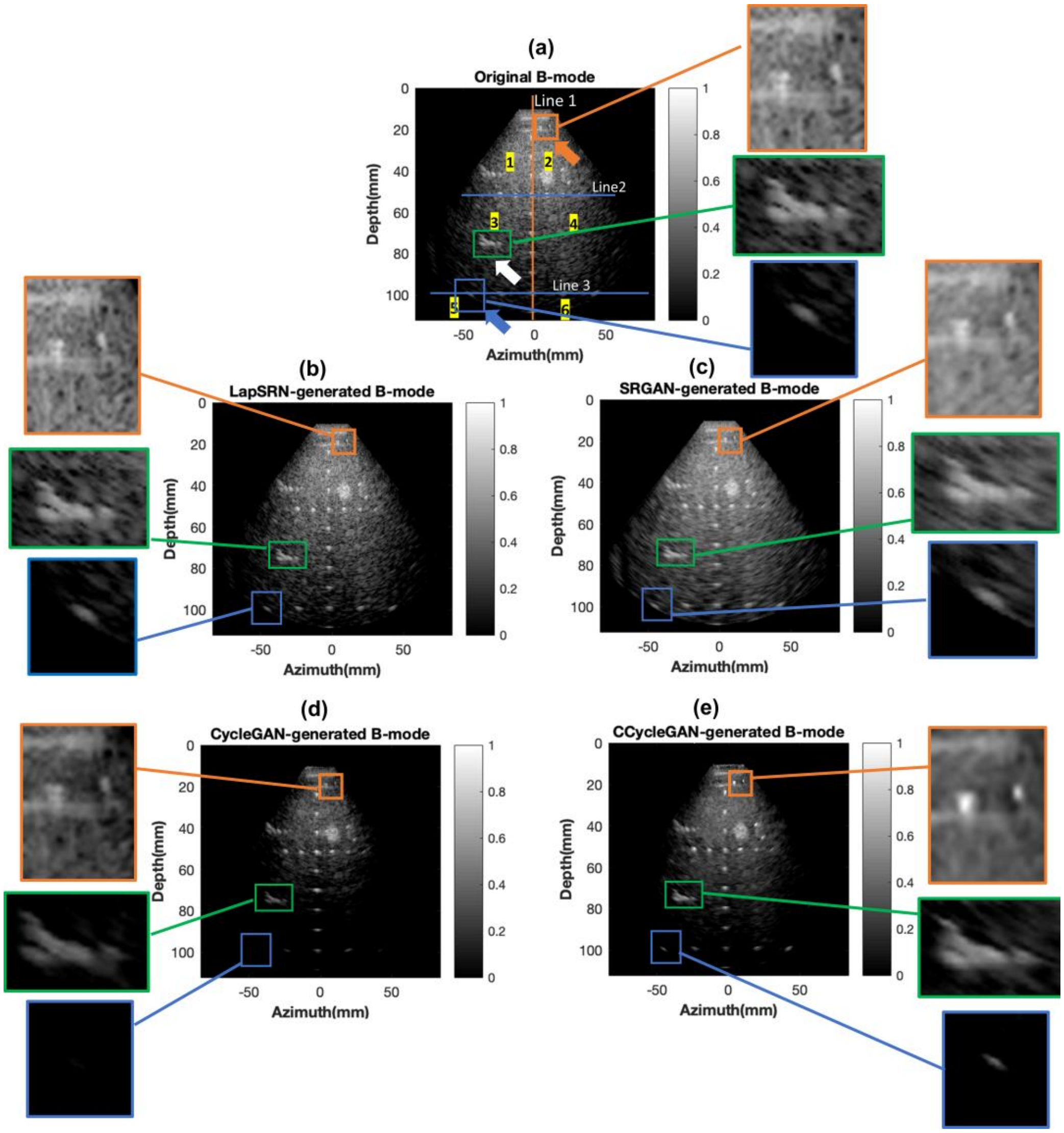

**Fig. 7.** Phased-array images in the full field-of-view and magnified regions of interest in View 2: **(a)** original image and model-generated B-mode images by **(b)** LapSRN, **(c)** SRGAN, **(d)** CycleGAN, and **(e)** CCycleGAN. The orange, green, and blue arrows indicate cropped regions for comparison.

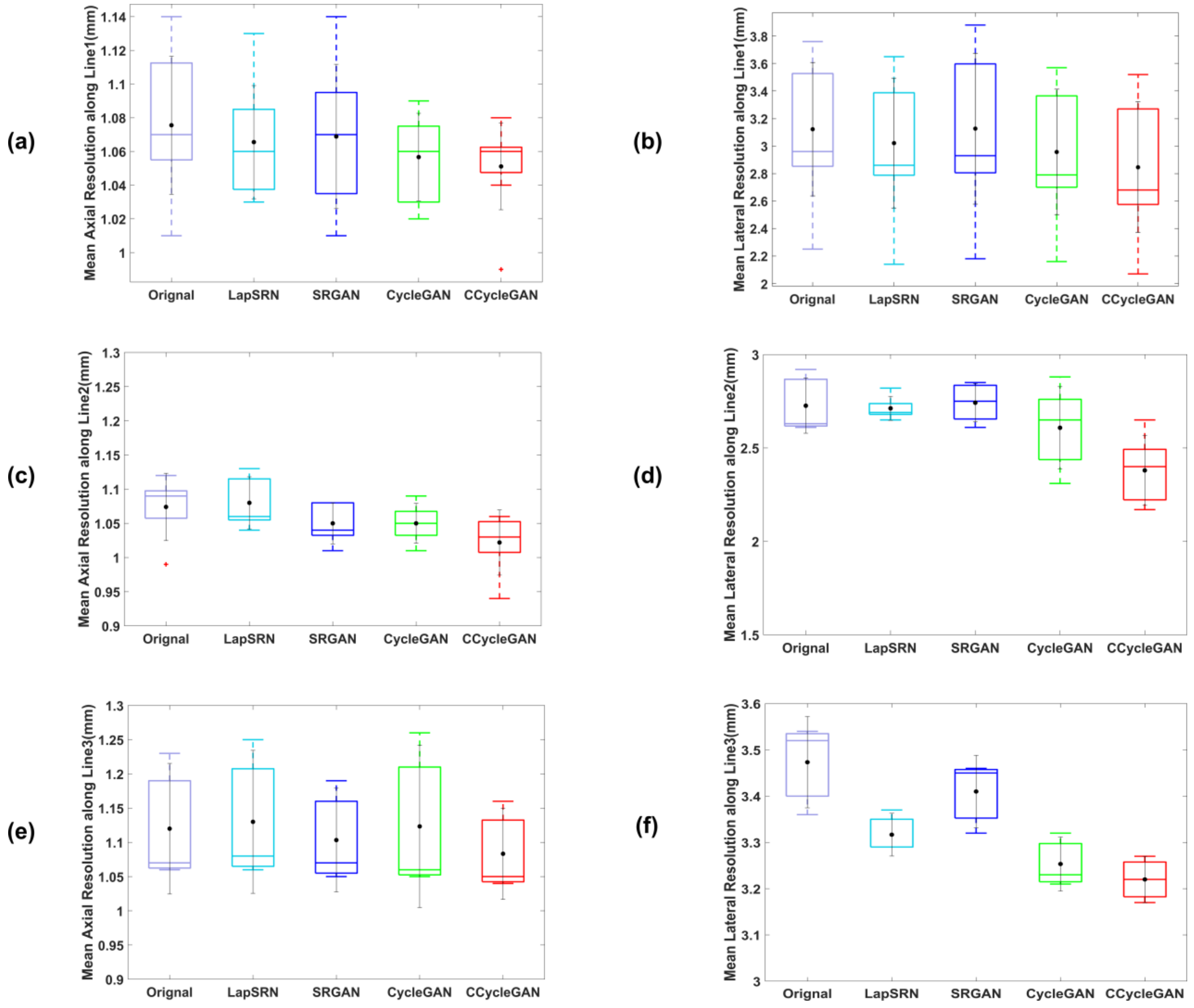

**Fig. 8** Mean axial resolution along **(a)** vertical Line 1, which includes Points 1-10, **(c)** horizontal Line 2, which contains Points 11-15, and **(e)** horizontal Line 3, where Points 16-18 lie. Mean lateral resolution along **(b)** vertical Line 1, **(d)** horizontal Line 2, and **(f)** horizontal Line 3. Note that the black dot in each box denotes the average value and that the black vertical line through each box represents the standard deviation (std) of the resolution along the corresponding line.

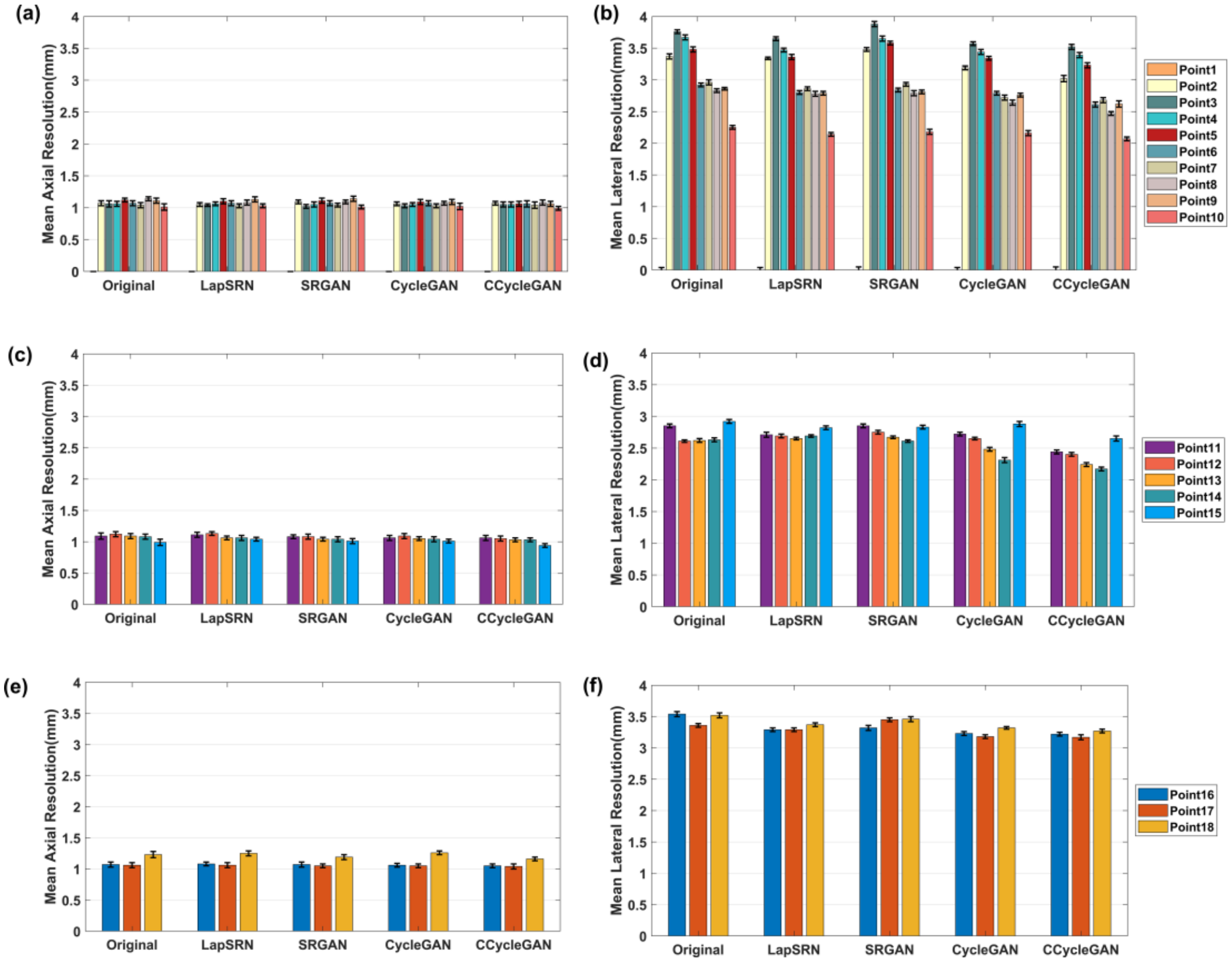

**Fig. 9.** The left panel shows bar charts of mean axial resolutions (mm) at point targets **(a)** 1-10, **(c)** 11-15, and **(e)** 16-18. The right panel shows bar charts of mean lateral resolutions (mm) at point targets **(b)** 1-10, **(d)** 11-15, and **(f)** 16-18. Error bars show standard deviations. Note that zero-mean values mean that concerned target points are incomplete.

Table 1. Statistical analysis of shape parameters *m* in **Region1**

| **Type** / **Mean±std** | **Original** | **LapSRN** | **SRGAN** | **CycleGAN** | **CCycleGAN** |
|---|---|---|---|---|---|
| **View 1** | 0.94±0.05 | 0.85±0.05 | 0.84±0.06 | 0.88±0.06 | 0.90±0.05 |
| **View 2** | 0.94±0.05 | 0.85±0.05 | 0.85±0.05 | 0.90±0.05 | 0.91±0.04 |
| **View 3** | 0.94±0.06 | 0.84±0.06 | 0.84±0.06 | 0.89±0.04 | 0.90±0.03 |

Table 2. Statistical analysis of shape parameters m in **Region2**

| **Type** / **Mean±td** | **Original** | **LapSRN** | **SRGAN** | **CycleGAN** | **CCycleGAN** |
|---|---|---|---|---|---|
| **View 1** | 0.94±0.06 | 0.84±0.04 | 0.85±0.05 | 0.88±0.05 | 0.90±0.04 |
| **View 2** | 0.94±0.05 | 0.85±0.05 | 0.85±0.05 | 0.88±0.05 | 0.90±0.05 |
| **View 3** | 0.94±0.05 | 0.84±0.05 | 0.84±0.05 | 0.89±0.06 | 0.90±0.05 |

Table 3. Statistical analysis of shape parameters m in **Region3**

| Type / Mean±std | Original | LapSRN | SRGAN | CycleGAN | CCycleGAN |
|---|---|---|---|---|---|
| **View 1** | 0.94±0.04 | 0.86±0.05 | 0.84±0.05 | 0.89±0.06 | 0.91±0.04 |
| **View 2** | 0.94±0.04 | 0.86±0.05 | 0.85±0.05 | 0.89±0.05 | 0.91±0.04 |
| **View 3** | 0.94±0.04 | 0.85±0.06 | 0.85±0.05 | 0.89±0.05 | 0.90±0.05 |

### *4.3 Speckle tracking*

**Fig. 10a** shows the cross-correlation (CC) coefficient and 2D (i.e., axial and lateral) displacement maps obtained from the original and deep learning model-generated images in the apical 2CH (**Fig. 10a**), apical 4CH, and parasternal SA views. The CC coefficient is a similarity measure for motion estimation of two successive ultrasound image frames. Semi-automatic segmentation of the left heart wall was further divided into two subregions, including left ventricular (LV) and left atrial (LA) walls, to evaluate the feasibility of speckle tracking on model-generated images. **Fig. 10b** shows a pre-deformed B-mode image (1st row) and the motion-corrected post-deformed B-mode image (2nd row) in the 2CH view. **Fig. 11** shows SSIM plots obtained from the entire left heart wall and two subregions in the 2CH view. Similar SSIM plots in 4CH and SA were obtained. The mean SSIM values from the entire left heart wall region in the case of CCycleGAN were slightly higher than those in benchmark cases (**Fig. 11** first row**)**. Considering the entire heart wall, CCycleGAN led to higher CC coefficients than LapSRN, SRGAN, and CycleGAN by 0.32%, 0.13% and 0.021%, respectively. Mean SSIM values of axial and lateral displacement maps obtained by CCycleGAN were higher than those by benchmarks. The average percent improvements of axial displacement estimation by CCycleGAN were respectively 2.28%, 0.76% and 0.13% compared with LapSRN, SRGAN, and CycleGAN. Meanwhile, CCycleGAN improved lateral displacement estimation by 2.06%, 0.92%, and 0.11%, respectively. As for region-based analyses (ROIs 1 and 2 in **Fig. 10a** ), mean SSIM values of both CC coefficients and 2D displacement maps from CCycleGAN were comparable to those from benchmarks (**Fig. 11**). CCycleGAN improved mean SSIM values of CC coefficients in ROIs 1 and 2 by 0.71%, 0.84%, and 0.13% compared with LapSRN, SRGAN, and CycleGAN, respectively. The average percent improvement of CC coefficients from CCycleGAN were respectively 0.07%, 0.03%, and 0.09%, compared with those from LapSRN, SRGAN, and CycleGAN.

**Table 4** summarized the RMSD values between pre-deformed and motion-corrected post-deformed images, excluding the background regions, under different image generation schemes. For overall speckle tracking accuracy throughout the entire heart wall, the smallest average RMSD value of 0.20 was from the CCycleGAN model output. The second smallest average RMSD value was 0.25 from the original B-mode data. In terms of regional speckle tracking accuracy, CCycleGAN also achieved the smallest RMSD values in the entire heart wall, ROI1 and ROI2. This demonstrated improvement of motion estimation from CCycleGAN-generated US images in the 2CH view compared with other model-generated images and original images.

### *4.4 Evaluation of SSIM and PSNR*

Both SSIM and PSNR between model-generated and original B-mode images of the *in vivo* human heart were calculated and shown in **Fig. 12**. Mean SSIM values from 2CH, 4CH, and SA views by CCycleGAN were 0.90, 0.91, and 0.91, respectively, and they were higher than those by LapSRN, SRGAN, and CycleGAN models. Similarly, mean PSNR values from 2CH, 4CH, and SA views in the case of CCycleGAN were 34.3, 35.9, and 36.8, respectively. This suggested that CCycleGAN-generated B-mode images resembled reference (i.e., original) B-mode images more than benchmarks. Overall, the statistics of SSIM and PSNR demonstrated that the proposed CCycleGAN improved the quality of cardiac US images.

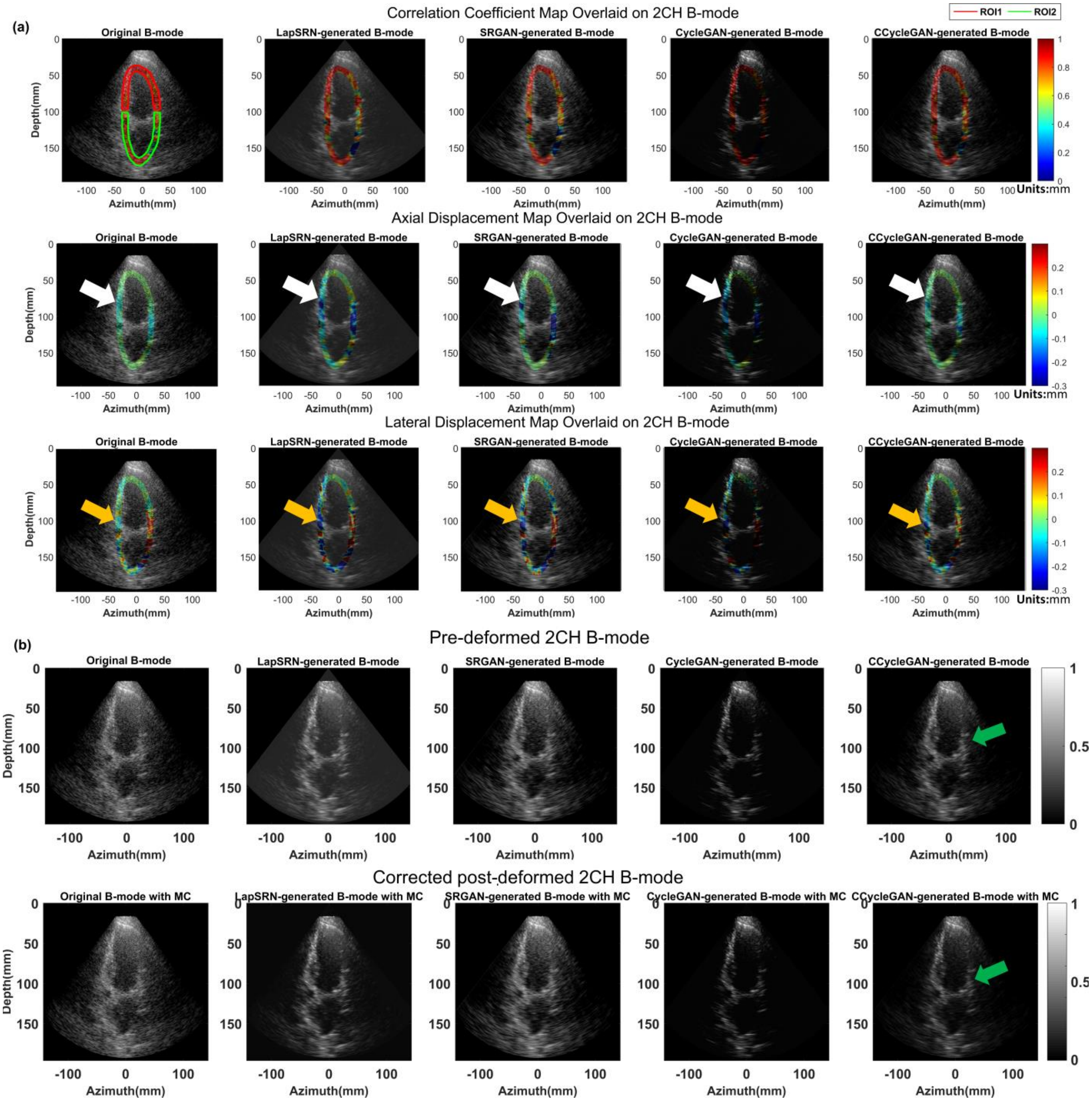

**Fig. 10.** (a) Exemplary correlation coefficient, inter-frame axial displacement, and inter-frame lateral displacement maps in the 2CH view. (b) Exemplary pre-deformed B-mode image and motion-corrected post-deformed image in the 2CH view. Note that the white, orange, and green arrows pinpoint exemplary myocardial segments exhibiting differences in tracking results.

Table 4. RMSD values between pre-deformed and motion-corrected post-deformed images in the entire heart wall, ROI 1, and ROI 2 in the 2CH view.

| Type / RMSD | Original | LapSRN | SRGAN | CycleGAN | CCycleGAN |
|---|---|---|---|---|---|
| **Entire heart wall** | 0.25±0.06 | 0.30±0.04 | 0.28±0.06 | 0.27±0.05 | **0.20±0.04** |
| **ROI 1** | 0.27±0.06 | 0.32±0.07 | 0.30±0.07 | 0.28±0.04 | **0.22±0.06** |
| **ROI 2** | 0.25±0.08 | 0.28±0.08 | 0.26±0.07 | 0.27±0.05 | **0.19±0.05** |

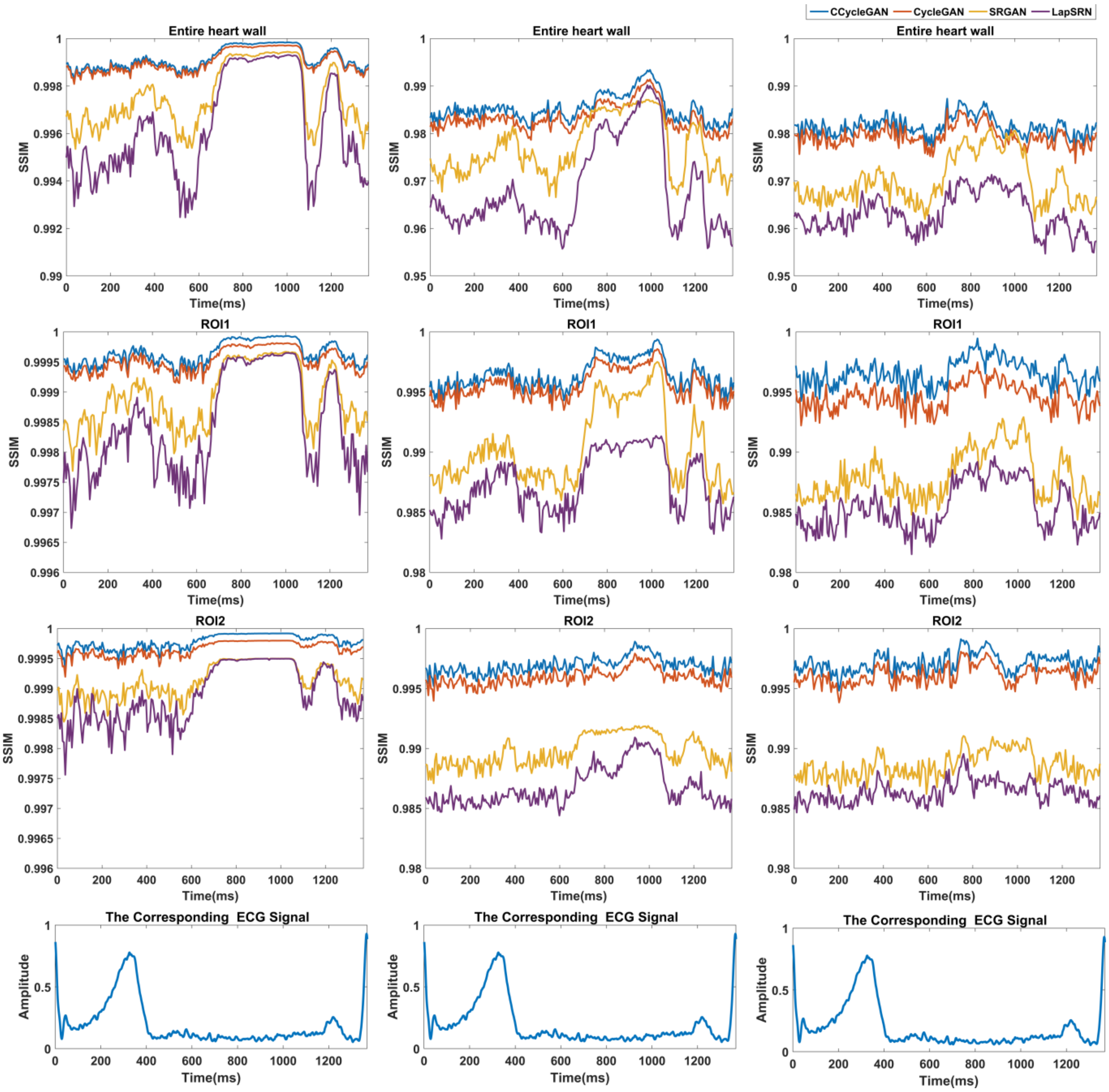

**Fig. 11.** Temporal profiles of SSIM values of correlation coefficient (left column), incremental axial displacement (middle column), and incremental lateral displacement (right column) maps obtained by speckle tracking of ultrasound 2CH images in one cardiac cycle. The top three rows are quantitative analyses from the entire heart wall, ROI1, and ROI2, respectively. The bottommost row shows the corresponding electrocardiography (ECG) signals of one cardiac cycle. Note that the first frame of the 2CH cine-loop data corresponds to the R-wave in the ECG signal.

### *4.5 Ablation study analysis*

**Table 5** summarized quantitative analyses of 1) the spatial resolution at **point target 16**, which was in a deep region, and 2) Nakagami $m$ parameters in the three regions (**Fig. 5b**) to examine the individual and combined effects of the proposed identical loss ($\mathcal{L}_{idt}$) and correlation coefficient loss ($\mathcal{L}_{cc}$) on the model performance. Comparing $\mathcal{L}_{idt}$, $\mathcal{L}_{cc}$, and the baseline (**Table 7** first row), we found that $\mathcal{L}_{idt}$ and $\mathcal{L}_{cc}$ improved CycleGAN significantly for generating images with improved spatial resolution and speckle pattern consistency. At the point target 16, $\mathcal{L}_{idt}$ + $\mathcal{L}_{cc}$ outperformed $\mathcal{L}_{idt}$ or $\mathcal{L}_{cc}$. This demonstrated that combining $\mathcal{L}_{idt}$ and $\mathcal{L}_{cc}$ with the conventional losses ($\mathcal{L}_{adv}$ and $\mathcal{L}_{cyc}$) in CycleGAN improved the model performance, particularly the image spatial resolution. According to Nakagami $m$ values calculated from **Regions 1-3** of the CycleGAN model-generated images in View 2 (**Fig. 5b**), $\mathcal{L}_{idt}$ did not affect speckle statistics, but $\mathcal{L}_{cc}$ resulted in $m$ values that were closer to those of the original input images (**Tables 1-3**). This verified that adding $\mathcal{L}_{cc}$ only into the final loss function enhanced the speckle pattern consistency. Results substantiated that the proposed CCycleGAN with $\mathcal{L}_{idt}$+$\mathcal{L}_{cc}$ outperformed the baseline model—CycleGAN without $\mathcal{L}_{idt}$ and $\mathcal{L}_{cc}$.

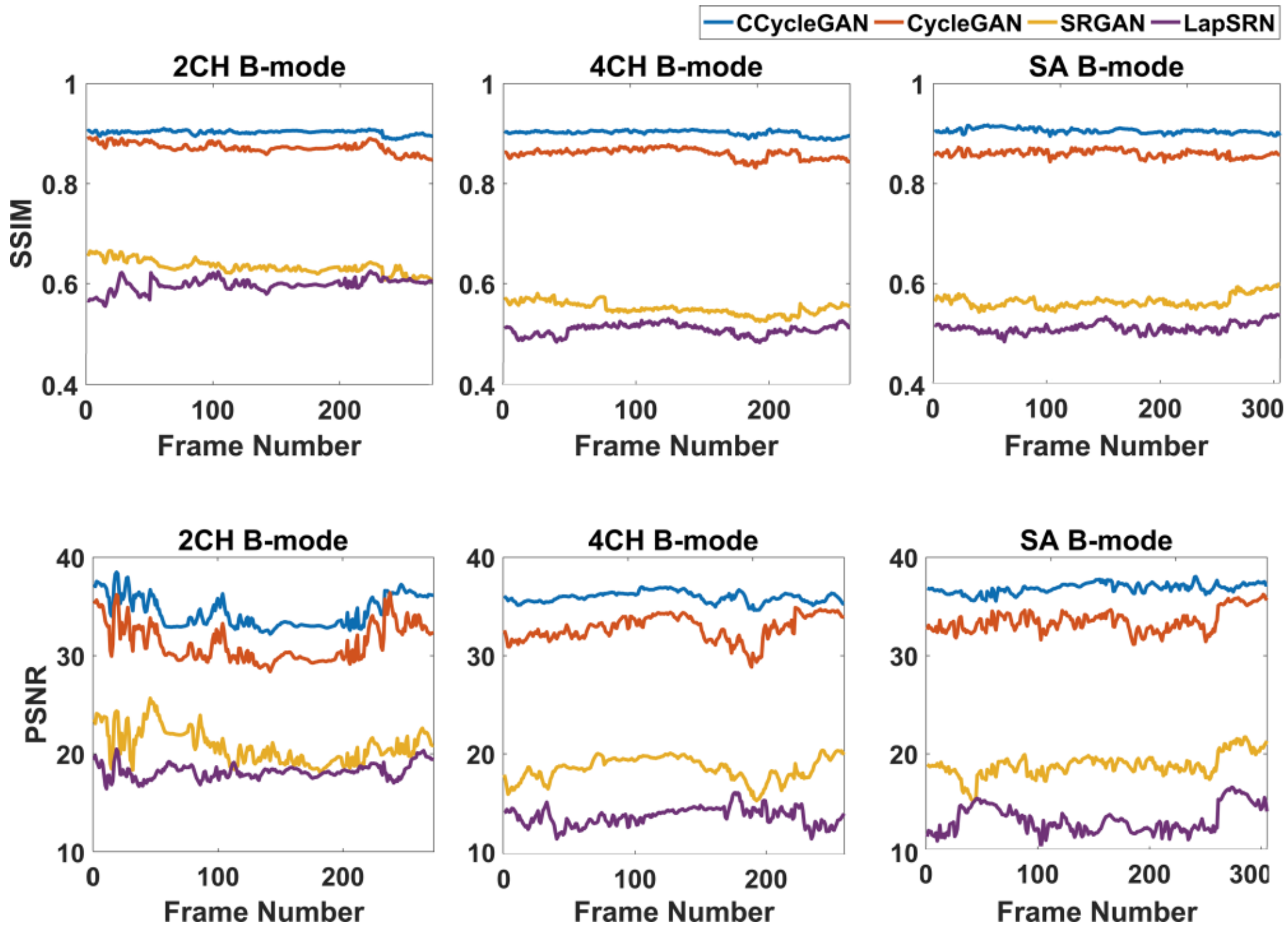

**Fig. 12.** Temporal profiles of SSIM and PSNR of cardiac B-mode images in three standard anatomical views by the proposed CCycleGAN and benchmark models—LapSRN, SRGAN, and CycleGAN.

Table 5. Comparison of quantitative results between the proposed CCycleGAN and conventional CycleGAN with or without $\mathcal{L}_{idt}$ or $\mathcal{L}_{cc}$. The average lateral (**16-lateral**) and axial (**16-Axial**) resolutions at a representative point target 16 and Nakagami $m$ values (***m*-region**) in three regions of model-generated images in View 2. The '√' symbol represents the inclusion of a concerned loss. The results from the proposed CCycleGAN model with identical and correlation coefficient losses are highlighted in bold.

| $\mathcal{L}_{idt}$ | $\mathcal{L}_{cc}$ | *16*-LATERAL | *16*-AXIAL | *M*-REGION1 | *M*-REGION2 | *M*-REGION3 |
|---|---|---|---|---|---|---|
| | | 3.54 ± 0.03 | 1.65 ± 0.03 | 0.90 ± 0.05 | 0.88 ± 0.05 | 0.89 ± 0.05 |
| √ | | 3.31 ± 0.02 | 1.81 ± 0.02 | 0.90 ± 0.05 | 0.88 ± 0.04 | 0.89 ± 0.04 |
| | √ | 3.32 ± 0.03 | 1.82 ± 0.03 | 0.91 ± 0.04 | 0.90 ± 0.05 | 0.91 ± 0.05 |
| √ | √ | **3.22 ± 0.04** | **1.64 ± 0.03** | **0.91 ± 0.04** | **0.90 ± 0.05** | **0.91 ± 0.04** |

## 5. Discussion

With the rapid development of promising data-to-data translation in the generative frameworks in recent years, few studies (Hu *et al.*, 2017; Wang *et al.*, 2019; Zhou *et al.*, 2020; Posilović *et al.*, 2022) started to focus on the improvement of ultrasound image quality by GAN or variations of GAN (Tom and Sheet, 2018; Lan *et al.*, 2019; He *et al.*, 2020; Posilović *et al.*, 2022). The compelling performance of GAN for ultrasound imaging has been demonstrated in various ultrasound imaging tasks, such as contrast improvement and noise reduction (Zhou *et al.*, 2020; Cronin *et al.*, 2020), beamforming (Goudarzi *et al.*, 2019; Nair *et al.*, 2019; Wang *et al.*, 2020; Zhou *et al.*, 2021), and image segmentation (Alsinan *et al.*, 2020; Han *et al.*, 2020). Nevertheless, the only sigmoid cross-entropy loss function often causes unstable training in the classical GAN. To address this issue, we were inspired to employ the losses of the CycleGAN (Zhu *et al.*, 2017) and further proposed two extra losses in CCycleGAN to investigate the scarcely explored issue—spatial resolution in sector images.

The main novelty of our CCycleGAN model lies in the addition of an identical loss and a correlation coefficient loss that are specific to ultrasound image characteristics to improve the spatial resolution in sector images. CCycleGAN, which take unpaired images, is more flexible than GAN variants, such as LapSRN and SRGAN, both of which use paired images. The results reported in Section 4 showed that CCycleGAN outperformed LapSRN, SRGAN, CycleGAN in the spatial resolution of the ultrasound images of an *in vitro* calibration phantom and the *in vivo* human heart. CCycleGAN generated B-mode images with better image quality than benchmarks as highlighted in **Regions 1-4** in **Fig. 7**. This was mainly because CCycleGAN introduced a correlation coefficient loss based on the structural similarity to preserve signal correlation of the point targets among consecutively generated US images. Particularly in the deep zones (e.g., **Regions 5-6** in **Fig. 7**), the CCycleGAN-generated images had better spatial resolutions than the CycleGAN-generated ones because of the added identical loss to generators that forced the entire tissue structure in the generated images to be indistinguishable from that in the input images. Similarly, in the SRGAN model, the loss functions, such as content and perceptual losses, constrained the trained model to generate high-quality US images with high SSIM and PSNR values. That was why the statistics of SSIM and PSNR values based on SRGAN were significantly higher than those based on LapSRN as shown in **Fig. 12**. In the analysis of speckle tracking, CCycleGAN also improved the quality of displacement estimates because higher US image quality ameliorated speckle tracking performance. The SSIM results in **Fig. 11** agreed well with CC coefficient maps in **Fig. 10**. The higher the CC coefficient, the less decorrelation between two consecutive ultrasound image frames and thus more reliable speckle tracking.

However, there are some limitations of the present CCycleGAN model, which trained *in vitro* phantom ultrasound images with fixed imaging parameters, such as the imaging depth and center frequency. Although the proposed CCycleGAN model showed better cardiac US image quality than the state-of-the-art deep learning models, an advanced framework is imperative to generalize the training datasets with various imaging parameters. Besides, our CCycleGAN model was validated *in vivo* in only one organ—the heart. How CCycleGAN performs in the improvement of ultrasound image quality of other human body parts should be examined. More extensive studies are warranted to further validate the generalizability of the proposed CCycleGAN model despite our promising preliminary results from *in vivo* cardiac ultrasound images.

The proposed CCycleGAN model incorporated two extra loss functions to preserve the intrinsic backscattering property of ultrasound imaging. Alternative loss functions may be considered; for example, within the same framework, we may extend other loss functions for the generator and the discriminator. With the recent advent of diffusion probabilistic models as a promising new generative model, the performance of our CCycleGAN framework may be optimized with the diffusion model concept to generate less spatially-varying US images. GANs can trade off image diversity for fidelity, thus producing high-quality images; however, it does not achieve robust training performance because of its partial consideration of the entire data distribution (Dhariwal and Nichol, 2021). Given diffusion models, our generative model in ultrasound imaging may even include more data domain properties (e.g. operating frequencies and speckle distributions) of US imaging to construct multi-scale learning strategies, such as multi-frequency learning (Batzolis *et al.*, 2022).

## 6. Conclusion

In this study, we proposed CCycleGAN, which uniquely consists of an identical and a correlation coefficient loss function pertinent to inherent ultrasound signal properties. It is a robust and efficient method for improving the spatial resolution of the sector US images of an *in vitro* multi-purpose multi-tissue phantom and the *in vivo* human beating heart by learning-based translation of unpaired phased-array to linear-array US images. The two proposed extra losses respectively constrained structural consistency and backscattering patterns between input and generated US images. Moreover, a concatenation layer and skip connections were employed to further improve the performance of the generator, and instance normalization with LeakyReLU was used to stabilize the discriminator. Five quality metrics, including spatial resolution, speckle statistics, RMSD associated with speckle tracking, SSIM, and PSNR, substantiated that CCycleGAN outperformed benchmarks and that the improved spatial resolution of CCycleGAN-generated cardiac US images led to better estimation quality of the heart wall motion.

## Acknowledgement

This study is in part supported by Midstream Research Programme for Universities under Innovation and Technology Commission (MRP/072/17X).

## References

Ahn H and Yim C 2020 Convolutional Neural Networks Using Skip Connections with Layer Groups for Super-Resolution Image Reconstruction Based on Deep Learning *Applied Sciences* **10** 1959

Alsinan A Z, Rule C, Vives M, Patel V M and Hacihaliloglu I Year *GAN-Based Realistic Bone Ultrasound Image and Label Synthesis for Improved Segmentation* Published *International Conference on Medical Image Computing and Computer-Assisted Intervention,2020)*, vol. Series*)*: Springer) pp 795-804

Batzolis G, Stanczuk J, Schönlieb C-B and Etmann C 2022 Non-Uniform Diffusion Models *arXiv preprint arXiv:2207.09786*

Cai Z, Xiong Z, Xu H, Wang P, Li W and Pan Y 2021 Generative adversarial networks: A survey toward private and secure applications *ACM Computing Surveys (CSUR)* **54** 1-38

Chartsias A, Joyce T, Dharmakumar R and Tsaftaris S A Year *Adversarial image synthesis for unpaired multi-modal cardiac data* Published *International Workshop on Simulation and Synthesis in Medical Imaging,2017)*, vol. Series*)*: Springer) pp 3-13

Cobbold R S 2006 *Foundations of biomedical ultrasound*: Oxford university press)

Cronin N J, Finni T and Seynnes O 2020 Using deep learning to generate synthetic B-mode musculoskeletal ultrasound images *Computer methods and programs in biomedicine* **196** 105583

D'hooge J, Heimdal A, Jamal F, Kukulski T, Bijnens B, Rademakers F, Hatle L, Suetens P and Sutherland G R 2000 Regional strain and strain rate measurements by cardiac ultrasound: principles, implementation and limitations *European Journal of Echocardiography* **1** 154-70

Dhariwal P and Nichol A 2021 Diffusion models beat gans on image synthesis *Advances in Neural Information Processing Systems* **34** 8780-94

Dietrichson F, Smistad E, Ostvik A and Lovstakken L Year *Ultrasound speckle reduction using generative adversial networks* Published *2018 IEEE International Ultrasonics Symposium (IUS),Mar 2018),* vol. Series*)*: IEEE) pp 1-4

Ding J, Zhao S, Tang F and Ning C Year *Ultrasound Image Super-Resolution with Two-Stage Zero-Shot CycleGAN* Published *Journal of Physics: Conference Series,2021),* vol. Series 2031*)*: IOP Publishing) p 012015

Fenster A and Downey D B 1996 3-D ultrasound imaging: A review *IEEE Engineering in Medicine and Biology magazine* **15** 41-51

Ge Y, Xue Z, Cao T and Liao S Year *Unpaired whole-body MR to CT synthesis with correlation coefficient constrained adversarial learning* Published *Medical Imaging 2019: Image Processing,2019),* vol. Series 10949*)*: International Society for Optics and Photonics) p 1094905

Goodfellow I, Pouget-Abadie J, Mirza M, Xu B, Warde-Farley D, Ozair S, Courville A and Bengio Y Year *Generative adversarial nets* Published *Advances in neural information processing systems,2014),* vol. Series*)* pp 2672-80

Goudarzi S, Asif A and Rivaz H Year *Multi-focus ultrasound imaging using generative adversarial networks* Published *2019 IEEE 16th international symposium on biomedical imaging (ISBI 2019),2019),* vol. Series*)*: IEEE) pp 1118-21

Han L, Huang Y, Dou H, Wang S, Ahamad S, Luo H, Liu Q, Fan J and Zhang J 2020 Semi-supervised segmentation of lesion from breast ultrasound images with attentional generative adversarial network *Comput. Methods Programs Biomed.* **189** 105275

He K, Zhang X, Ren S and Sun J Year *Deep residual learning for image recognition* Published *Proceedings of the IEEE conference on computer vision and pattern recognition,2016),* vol. Series*)* pp 770-8

He X, Lei Y, Liu Y, Tian Z, Wang T, Curran W, Liu T and Yang X Year *Deep attentional GAN-based high-resolution ultrasound imaging* Published *Medical Imaging 2020: Ultrasonic Imaging and Tomography.International Society for Optics and Photonics,2020),* vol. Series 11319*)*: SPIE)

Ho J, Jain A and Abbeel P 2020 Denoising diffusion probabilistic models *Advances in Neural Information Processing Systems* **33** 6840-51

Hu Y, Gibson E, Lee L-L, Xie W, Barratt D C, Vercauteren T and Noble J A Year *Freehand Ultrasound Image Simulation with Spatially-Conditioned Generative Adversarial Networks* Published *(Cham, 2017) (Molecular Imaging, Reconstruction and Analysis of Moving Body Organs, and Stroke Imaging and Treatment,* vol. Series*)*: Springer International Publishing) pp 105-15

Isola P, Zhu J-Y, Zhou T and Efros A A Year *Image-to-image translation with conditional adversarial networks* Published *Proceedings of the IEEE conference on computer vision and pattern recognition,2017),* vol. Series*)* pp 1125-34

Jensen J A, Nikolov S I, Gammelmark K L and Pedersen M H 2006 Synthetic aperture ultrasound imaging *Ultrasonics* **44** e5-e15

Johnson J, Alahi A and Fei-Fei L Year *Perceptual losses for real-time style transfer and super-resolution* Published *European conference on computer vision,2016),* vol. Series*)*: Springer) pp 694-711

Kang E, Koo H J, Yang D H, Seo J B and Ye J C 2019 Cycle-consistent adversarial denoising network for multiphase coronary CT angiography *Medical physics* **46** 550-62

Kingma D P and Ba J 2014 Adam: A method for stochastic optimization *arXiv preprint arXiv:1412.6980*

Lan H, Zhou K, Yang C, Cheng J, Liu J, Gao S and Gao F Year *Ki-GAN: Knowledge Infusion Generative Adversarial Network for Photoacoustic Image Reconstruction In Vivo* Published *International Conference on Medical Image Computing and Computer-Assisted Intervention,2019) (Medical Image*

*Computing and Computer Assisted Intervention – MICCAI 2019,* vol. Series*)*: Springer International Publishing) pp 273-81

Ledig C, Theis L, Huszár F, Caballero J, Cunningham A, Acosta A, Aitken A, Tejani A, Totz J and Wang Z Year *Photo-realistic single image super-resolution using a generative adversarial network* Published *Proceedings of the IEEE conference on computer vision and pattern recognition,2017),* vol. Series*)* pp 4681-90

Li H, Guo Y and Lee W-N 2016 Systematic performance evaluation of a cross-correlation-based ultrasound strain imaging method *Ultrasound in medicine & biology* **42** 2436-56

Li H and Lee W-N 2017 Effects of tissue mechanical and acoustic anisotropies on the performance of a cross-correlation-based ultrasound strain imaging method *Physics in Medicine & Biology* **62** 1456

Liebgott H, Rodriguez-Molares A, Cervenansky F, Jensen J A and Bernard O Year *Plane-wave imaging challenge in medical ultrasound* Published *2016 IEEE International Ultrasonics Symposium (IUS),2016),* vol. Series*)*: IEEE) pp 1-4

Liu H, Liu J, Hou S, Tao T and Han J 2021 Perception consistency ultrasound image super-resolution via self-supervised CycleGAN *Neural Computing and Applications* 1-11

Martin Arjovsky S and Bottou L Year *Wasserstein generative adversarial networks* Published *Proceedings of the 34 th International Conference on Machine Learning,2017),* vol. Series*)* pp 214-23

Mishra D, Chaudhury S, Sarkar M and Soin A S 2018 Ultrasound image enhancement using structure oriented adversarial network *IEEE Signal Processing Letters* **25** 1349-53

Mondillo S, Galderisi M, Mele D, Cameli M, Lomoriello V S, Zacà V, Ballo P, D'Andrea A, Muraru D and Losi M 2011 Speckle-tracking echocardiography: a new technique for assessing myocardial function *Journal of Ultrasound in Medicine* **30** 71-83

Montaldo G, Tanter M, Bercoff J, Benech N and Fink M 2009 Coherent plane-wave compounding for very high frame rate ultrasonography and transient elastography *IEEE transactions on ultrasonics, ferroelectrics, and frequency control* **56** 489-506

Nair A A, Tran T D, Reiter A and Bell M A L Year *A generative adversarial neural network for beamforming ultrasound images: Invited presentation* Published *2019 53rd Annual conference on information sciences and systems (CISS),2019),* vol. Series*)*: IEEE) pp 1-6

Nehra R, Pal A and Baranidharan B Year *Radiological Image Synthesis Using Cycle-Consistent Generative Adversarial Network* Published *Proceedings of International Conference on Recent Trends in Computing,2022),* vol. Series*)*: Springer) pp 391-402

Papadacci C, Pernot M, Couade M, Fink M and Tanter M 2014 High-contrast ultrafast imaging of the heart *IEEE transactions on ultrasonics, ferroelectrics, and frequency control* **61** 288-301

Posilović L, Medak D, Subašić M, Budimir M and Lončarić S 2022 Generating ultrasonic images indistinguishable from real images using Generative Adversarial Networks *Ultrasonics* **119** 106610

Radford A, Metz L and Chintala S 2015 Unsupervised representation learning with deep convolutional generative adversarial networks *arXiv preprint arXiv:1511.06434*

Roy S, Butman J A and Pham D L Year *Synthesizing CT from ultrashort echo-time MR images via convolutional neural networks* Published *International Workshop on Simulation and Synthesis in Medical Imaging,2017),* vol. Series*)*: Springer) pp 24-32

Royer D F 2019 Seeing with sound: how ultrasound is changing the way we look at anatomy *Biomedical Visualisation: Volume 2* 47-56

Schaefferkoetter J, Yan J, Moon S, Chan R, Ortega C, Metser U, Berlin A and Veit-Haibach P 2021 Deep learning for whole-body medical image generation *European journal of nuclear medicine and molecular imaging* 1-10

Schapire R E, Freund Y, Bartlett P and Lee W S 1998 Boosting the margin: A new explanation for the effectiveness of voting methods *The annals of statistics* **26** 1651-86

Tai Y, Yang J and Liu X Year *Image super-resolution via deep recursive residual network* Published *Proceedings of the IEEE conference on computer vision and pattern recognition,2017),* vol. Series*)* pp 3147-55

Tom F and Sheet D Year *Simulating patho-realistic ultrasound images using deep generative networks with adversarial learning* Published *2018 IEEE 15th International Symposium on Biomedical Imaging (ISBI 2018),Apr 2018),* vol. Series*)* pp 1174-7

Ulyanov D, Vedaldi A and Lempitsky V 2016 Instance normalization: The missing ingredient for fast stylization *arXiv preprint arXiv:1607.08022*

Wang C, Xu C, Wang C and Tao D 2018 Perceptual adversarial networks for image-to-image transformation *IEEE Transactions on Image Processing* **27** 4066-79

Wang R, Fang Z, Gu J, Guo Y, Zhou S, Wang Y, Chang C and Yu J 2019 High-resolution image reconstruction for portable ultrasound imaging devices *EURASIP Journal on Advances in Signal Processing* **2019** 1-12

Wang Y, Kempski K, Kang J U and Bell M A L Year *A Conditional Adversarial Network for Single Plane Wave Beamforming* Published *2020 IEEE International Ultrasonics Symposium (IUS),2020),* vol. Series*)*: IEEE) pp 1-4

Wolterink J M, Dinkla A M, Savenije M H, Seevinck P R, van den Berg C A and Išgum I Year *Deep MR to CT synthesis using unpaired data* Published *International Workshop on Simulation and Synthesis in Medical Imaging,Sep 2017),* vol. Series*)*: Springer) pp 14-23

Yang H, Sun J, Carass A, Zhao C, Lee J, Xu Z and Prince J Year *Unpaired Brain MR-to-CT Synthesis Using a Structure-Constrained CycleGAN* Published *Deep Learning in Medical Image Analysis and Multimodal Learning for Clinical Decision Support,2018),* vol. Series*)* ed D Stoyanov*, et al.*: Springer International Publishing) pp 174-82

Yang L, Zhang Z, Song Y, Hong S, Xu R, Zhao Y, Shao Y, Zhang W, Cui B and Yang M-H 2022 Diffusion models: A comprehensive survey of methods and applications *arXiv preprint arXiv:2209.00796*

Yi J, Kang H K, Kwon J-H, Kim K-S, Park M H, Seong Y K, Kim D W, Ahn B, Ha K and Lee J 2021 Technology trends and applications of deep learning in ultrasonography: image quality enhancement, diagnostic support, and improving workflow efficiency *Ultrasonography* **40** 7

Yu X, Guo Y, Huang S-M, Li M-L and Lee W-N 2015 Beamforming effects on generalized Nakagami imaging *Physics in Medicine & Biology* **60** 7513

Yuan Y, Liu S, Zhang J, Zhang Y, Dong C and Lin L Year *Unsupervised image super-resolution using cycle-in-cycle generative adversarial networks* Published *Proceedings of the IEEE Conference on Computer Vision and Pattern Recognition Workshops,2018),* vol. Series*)* pp 701-10

Zhang Y, Guo Y and Lee W-N 2017 Ultrafast ultrasound imaging with cascaded dual-polarity waves *IEEE Transactions on Medical Imaging* **37** 906-17

Zhang Y, Li H and Lee W-N 2019 Imaging Heart Dynamics With Ultrafast Cascaded-Wave Ultrasound *IEEE transactions on ultrasonics, ferroelectrics, and frequency control* **66** 1465-79

Zhang Z, Yang L and Zheng Y Year *Translating and segmenting multimodal medical volumes with cycle- and shape-consistency generative adversarial network* Published *Proceedings of the IEEE conference on computer vision and pattern recognition,2018),* vol. Series*)* pp 9242-51

Zhou Z, Guo Y and Wang Y 2021 Ultrasound deep beamforming using a multiconstrained hybrid generative adversarial network *Medical Image Analysis* **71** 102086

Zhou Z, Wang Y, Guo Y, Qi Y and Yu J 2020 Image Quality Improvement of Hand-Held Ultrasound Devices With a Two-Stage Generative Adversarial Network *IEEE Transactions on Biomedical Engineering* **67** 298-311

Zhu J-Y, Park T, Isola P and Efros A A Year *Unpaired image-to-image translation using cycle-consistent adversarial networks* Published *Proceedings of the IEEE international conference on computer vision,2017),* vol. Series*)* pp 2223-32